\documentclass[12pt]{article}
\usepackage{amsfonts,amsthm,amsmath,amssymb,upgreek,bm}
\usepackage{hyperref}    % To enable hyperlinks
\hypersetup{
    colorlinks=true,     % Enable colored links
    linkcolor=blue,      % Color of internal links
    urlcolor=blue,       % Color of external URLs
    citecolor=blue       % Color of citations
}
\usepackage[paper=letterpaper,margin=.85in]{geometry}

\usepackage{units}
\usepackage{float}
\usepackage[export]{adjustbox}
\usepackage[shortlabels]{enumitem}
\usepackage[mathscr]{euscript}
\usepackage[normalem]{ulem}
\usepackage{soul}
\usepackage[T1]{fontenc} % if needed
\usepackage{array}
\usepackage{makecell}
\usepackage{amsfonts, bm, bbm}
\usepackage{mathrsfs}

\newcommand{\bra}[1]{\left\langle #1 \right|}
\newcommand{\ket}[1]{\left| #1 \right\rangle}
\newcommand{\pbra}[1]{\left( #1 \right|}
\newcommand{\pket}[1]{\left| #1 \right)}
\newcommand{\bracket}[2]{\langle #1 | #2 \rangle}
\newcommand{\tr}{\operatorname{tr}}
\newcommand{\lind}{\mathcal{L}}

\usepackage[dvipsnames]{xcolor}

\begin{document}
\thispagestyle{empty}

\vspace*{2.5cm}
\begin{center}

{\bf {\LARGE Measurement induced scrambling and emergent symmetries in random circuits}}

\begin{center}

\vspace{1cm}

{\bf Haifeng Tang$^1$, Hong-Yi Wang$^2$, Zhong Wang$^3$, Xiao-Liang Qi$^1$}\\
  \bigskip \rm

\bigskip 
${}^1$Stanford Institute for Theoretical Physics, Stanford University, Stanford, CA 94305, USA

${}^2$Princeton Quantum Initiative, Princeton University, Princeton, NJ 08544, USA

${}^3$Institute for Advanced Study, Tsinghua University, Beijing 100084, China

\rm
  \end{center}

\vspace{2cm}
{\bf Abstract}
\end{center}

%% %simple case: 2 authors, same institution
%% \author{A. Uthor}
%% \author{and A. Nother Author}
%% \affiliation{Institution,\\Address, Country}

% more complex case: 4 authors, 3 institutions, 2 footnotes

% The "\note" macro will give a warning: "Ignoring empty anchor\dots"
% you can safely ignore it.

% e-mail addresses: one for each author, in the same order as the authors

\begin{quotation}
    
\noindent
Quantum entanglement is affected by unitary evolution, which spreads the entanglement through the whole system, and also by measurements, which usually tends to disentangle subsystems from the rest. Their competition has been known to result in the measurement-induced phase transition. But more intriguingly, measurement alone has the ability to drive a system into different entanglement phases. In this work, we map the entanglement evolution under unitaries and/or measurements into a classical spin problem. This framework is used to understand a myriad of random circuit models analytically, including measurement-induced and measurement-only transitions. Regarding many-body joint measurements, a lower bound of measurement range that is necessary for a global scrambled phase is derived. Moreover, emergent continuous symmetries (U(1) or SU(2)) are discovered in some random measurement models in the large-$d$ (qudit dimension) limit. % and they are accompanied by a critical purification time. 
The emergent continuous symmetry allows a variety of spin dynamics phenomena to find their counterparts in random circuit models. \end{quotation}

\renewcommand{\thefootnote}{} % suppress number
%\footnotetext{hftang@stanford.edu, hywang@princeton.edu, wangzhongemail@mail.tsinghua.edu.cn, xlqi@stanford.edu}

\footnotetext{hftang@stanford.edu}
\footnotetext{hywang@princeton.edu}
\footnotetext{ wangzhongemail@mail.tsinghua.edu.cn}
\footnotetext{xlqi@stanford.edu}
\renewcommand{\thefootnote}{\arabic{footnote}} % restore default numbering

\setcounter{page}{0}
\setcounter{tocdepth}{2}
\setcounter{footnote}{0}
\newpage

\setcounter{page}{2}

\hypersetup{linkcolor=black}
\tableofcontents
\hypersetup{linkcolor=blue}

\section{Introduction}
% \setlength{\parskip}{0em}

%Quantum entanglement is an evergreen topic in both quantum gravity and condensed matter physics, through which we see the pronounced difference between quantum and classical world. Of most generality are the models in which quantum information spreads and scrambles very rapidly, which include most non-integrable many-body systems and supposedly the black hole dynamics. Hence a systematic understanding of information scrambling dynamics is desirable. 

%Especially, it is customary to model the ubiquitous highly scrambling dynamics with random circuit models \cite{RQC1,RQC2,RQC3,RQC4,RQC5}. The past few years have witnessed a surge in the area of random quantum circuits, since they are tamed to analytic study. In general, these models contain a union of qudits (microscopic $d$-level quantum systems), and they are subject to either local random unitary operators in discrete time, or time-dependent random Hamiltonian evolution. Averaging over the possible realizations of the evolution, the moments of the density operator (including entanglement entropies, out-of-time ordered correlators (OTOC), etc.) turn out to satisfy the same evolution as some spin many-body system. For a review of this methodology we recommend Ref.~\cite{RQC6}.

Quantum entanglement is a perennial topic in both quantum gravity and condensed matter physics, highlighting the pronounced difference between the quantum and classical worlds. Of particular generality are the models in which quantum information spreads and scrambles rapidly, encompassing most non-integrable many-body systems and, presumably, black hole dynamics. Consequently, a comprehensive understanding of information scrambling dynamics is desirable.

A common model for highly scrambling dynamics is random quantum circuit
\cite{RQC1,RQC2,RQC3,RQC4,RQC5}. The past few years have seen a surge in the area of random quantum circuits, as they have become more amenable to analytic study. Generally, these models consist of a collection of qudits (microscopic $d$-level quantum systems) that are subject to either local random unitary operators in discrete time or time-dependent random Hamiltonian evolution. By averaging over the possible realizations of the evolution, the moments of the density operator (including entanglement entropies, out-of-time ordered correlators (OTOC), etc.) are found to satisfy the same evolution as certain spin many-body systems. For a review of this methodology, we recommend Ref.~\cite{RQC6}.

Even more interestingly, the mapping between entanglement dynamics and spin models readily incorporates measurements. Typically, a projective measurement $\rho \to \ket{\varphi}\bra{\varphi} \rho \ket{\varphi}\bra{\varphi}$ may occur on a single qudit according to the Born probability at each instant of time, where $\ket{\varphi}$ represents a randomly chosen state. This measurement protocol results in a pure state, effectively disentangling the qudit from the rest of the system. Unlike unitary gates which usually increase entanglement of a system, measurements tend to purify, i.e., disentangle the system. Consequently, they compete and give rise to the well-known measurement-induced phase transition (MIPT) \cite{MIPT1,MIPT2,MIPT3,MIPT4,MIPT5,MIPT6,MIPT7,MIPT8,MIPT9,MIPT10,MIPT11,MIPT12,MIPT13,MIPT14,MIPT15,MIPT16,MIPT17,MIPT18,MIPT19,MIPT20,MIPT21,MIPT22,MIPT23,MIPT24,MIPT25,MIPT27,MIPT28,MIPT29,MIPT30,MIPT31}. In addition to entanglement entropy, other indicators of MIPT have been studied, such as mutual information \cite{bao2021teleportation}, entanglement negativity \cite{PRXQuantum.2.030313, MIPT22}, and cross entropy \cite{li2022cross}. Recently, MIPT has been experimentally observed on quantum computing devices \cite{noel2022measurement, koh2022experimental, google2023quantum}.

%Intuitively, with more unitary evolution, the system evolves into a highly scrambled final state exhibiting volume-law entropy; in contrast, with more measurement, the final state is nearly pure and displays area-law entropy. %The transition between the scrambled and purified phases can be either abrupt or smooth. 
When more generic measurements acting on multiple qudits are considered, the effects of measurements turn out to be to be far more versatile than just purification. Even in a measurement-only system with trivial unitary evolution, several types of phase transitions can occur by adjusting parameters \cite{PhysRevB.102.094204, PhysRevX.11.011030, vijay2022monitored}. Compared with single qudit projective measurements, multi-qudit measurements do not commute with each other, which is the key reason of ``measurement frustration" that induces scrambling through measurements.

%Obviously,if all the measurements applied commute with each other, the final state is inevitably purified. This is the case in Secs.~\ref{ssec:z2transition} and \ref{ssec:page}, where single-body measurement showing only the power of purification. Conversely, when some measurements do not commute with each other, a subsystem purified by a measurement may re-entangle with the rest of the system by subsequent measurements. 

Indeed, in reference \cite{PhysRevX.11.011030}, the authors applied certain classes of measurement-only circuits  on 1d/2d lattice spin chain model, and find numerical evidence that certain level of measurement ``frustration''(subsequent measurements being non-commuting) is necessary to generate a globally scrambled phase using only measurement. However, an analytical characterization of the conditions under which the measurement-only circuit would lead to scrambling instead of purification has not been fully developed. One purpose of our paper is to fill this gap. In a simpler setting by considering all-to-all connected lattice, the permutation symmetry aids us to analytical results. In recent years,  measurement induced correlation and entanglement has also been studied in~\cite{PhysRevA.110.032426}.

\begin{figure}[t]
    \centering
    \includegraphics[width=1\textwidth]{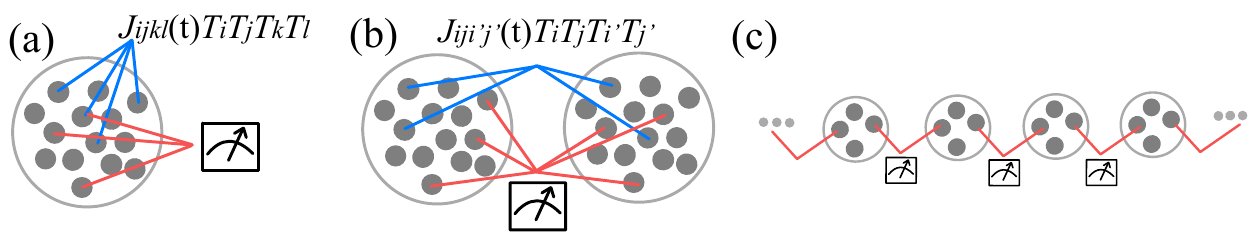}
    \caption{\label{fig:cluster illustration} Basic ingredients in the random circuit models we will study. (a) Brownian many-body Hamiltonian (blue) and random many-body projective measurement (red), see Sec.~\ref{sec:setup}. The demonstrated case is $q=4$-body unitary and $q'=3$-body measurement. (b) Two coupled clusters, cf. Sec.~\ref{ssec:2interaction}. We may include inter-cluster Brownian Hamiltonian evolution, or inter-cluster joint measurement. (c) One-dimensional chain of clusters with nearest-neighbor $(1,1)$-body joint measurement, see Sec.~\ref{ssec:su2}. }
\end{figure}

In this paper, we systematically investigate in the effect of measurements on quantum entanglement in a family of randomly coupled qudit models.  The setup of the model we study is sketched in Fig.~\ref{fig:cluster illustration}. We consider a lattice model with each site consisting of $N$ qudits. We consider Brownian $q$-body unitary coupling between these qudits, and random $q'$-body measurements. The unitary gates and the random measurements both include intra-site terms and inter-site terms.  One of the advantages of our model is that its purity evolution reduces to imaginary time evolution of familiar quantum spin models, which enables a new analytically controllable limit. %entanglement phase transition for purity can be studied analytically.

More specifically, we show that the averaged purity dynamics of this model can be mapped to a spin chain with $N+1$ states on each site, which can be conveniently viewed as a $\mathrm{SU}(2)$ spin with total angular momentum $N/2$. Entanglement phase transition thus corresponds to ground state phase transition in this spin model. For example, a volume law phase corresponds to a discrete spontaneous symmetry breaking phase, whereas an area law phase corresponds to a paramagnetic phase. Using preexisting knowledge about spin models, we are able to obtain analytic results on MIPT in this model. 
%, we establish a mathematical formalism to map the large-cluster entanglement evolution under hybrid random circuit to classical spin motion. The resulted classical problem is highly tractable. Our new formalism provides an understanding of entanglement phase transitions in terms of classical ground-state spontaneous symmetry breaking. But more importantly, we are able to diagnose what extent of measurement frustration is necessary for global scrambling. 
For models driven only by all-to-all measurements, there is an explicit relation between the qudit dimension $d$ and the range of measurement
\begin{equation}
    d_c = \frac{q'}{q'-2},
\end{equation}
which is the phase boundary between scrambled phase and purified phase. In particular, in the case of qubits ($d=2$), the critical point is at $q'=4$, which means $q'>4$-body all-to-all connected measurements lead to a volume law phase.% we need at least five-body measurement, suggesting the feasibility of measurement-induced scrambling but with relatively high cost. 

Another outcome of our formalism are different classes of emergent symmetry in measurement-only models. %Adhering to the original highly chaotic dynamics, these emergent symmetries would be difficult to discern, which is why this aspect has been overlooked previously. However, they can be easily extracted from the mapped spin model.
In the large $d$ limit, a $\mathrm{U}(1)$ symmetry emerges with two-body all-to-all projective measurements; while an $\mathrm{SU}(2)$ symmetry emerges with inter-cluster $(1,1)$-body measurements. In both cases, the symmetry group is larger than the apparent $\mathbb{Z}_2$, which is the subsystem inversion symmetry\footnote{Subsystem inversion symmetry (discussed in detail in Sec.~\ref{ssec:z2transition}) originates from the fact that each subsystem has the same entanglement entropy as its complement if the entire system is in a pure state. Purity of a state is preserved by both the unitary evolution and projective measurements. %Our construction of both the Hamiltonian evolution and projective measurements preserves a pure state's purity.
}. The significance of the emergent $\mathrm{SU}(2)$ symmetry we discovered is that it may inspire even more new entanglement phases, owing to the vast repertoire of spin many-body physics. For instance, in a one-dimensional (1D) cluster chain, we can map the entanglement dynamics to the imaginary-time evolution of an antiferromagnetic Heisenberg model. From the latter, for odd $N$, we predict a  CFT$_2$-like entanglement entropy in the steady state, and a polynomially long time is required to saturate the volume law due to the gaplessness. We refer to this phenomenon as critical purification. 

%The remainder of this paper is arranged as follows. In Sec.~\ref{sec:setup}, we establish the formalism that maps the entanglement entropy dynamics into a large-spin Hamiltonian evolution in imaginary time. With the mapped Hamiltonian, we are able to analyze entanglement phase transition in all-to-all models in Sec.~\ref{sec:alltoall}. We obtain an analytic expression for the entanglement entropy in the scrambled phase as a Page curve in Sec.~\ref{ssec:page}. We particularly concentrate on measurement-only dynamics in Sec.~\ref{ssec:MIS}, in which we analytically obtained the phase boundary between measurement-induced scrambling and measurement-induced purification, along with an emergent $\mathrm{U}(1)$ symmetry in large $d$. In Sec.~\ref{sec:spatial}, we generalize our argument to random quantum circuits with spatial locality (instead of all-to-all interaction), and find that enlarged symmetry group is possible. Then we discuss in detail the consequences of emergent $\mathrm{SU}(2)$ symmetry in inter-cluster measurement model (Sec.~\ref{ssec:su2}). We make final comments and address unsolved questions in Sec.~\ref{sec:conclusion}. 

The remainder of this paper is organized as follows. In Sec.\ref{sec:setup}, we develop the formalism that maps the entanglement entropy dynamics to a large-spin Hamiltonian evolution in imaginary time. Using the mapped Hamiltonian, we analyze entanglement phase transitions in all-to-all models in Sec.\ref{sec:alltoall}. We derive an analytic expression for the entanglement entropy in the scrambled phase as a Page curve in Sec.\ref{ssec:page}. We specifically focus on measurement-only dynamics in Sec.\ref{ssec:MIS}, where we analytically obtain the phase boundary between measurement-induced scrambling and measurement-induced purification, along with an emergent $\mathrm{U}(1)$ symmetry for large $d$. In Sec.\ref{sec:spatial}, we extend our argument to random quantum circuits with spatial locality (rather than all-to-all interaction) and find that an enlarged symmetry group is possible. We then discuss in detail the implications of the emergent $\mathrm{SU}(2)$ symmetry in the inter-cluster measurement model (Sec.\ref{ssec:su2}). We conclude with final remarks and address open questions in Sec.~\ref{sec:conclusion}.

\section{\label{sec:setup}General setup}

\subsection{\label{ssec:map_classical}From random circuits to large spins}

We begin with the general methodology of studying the evolution of entanglement entropy in random quantum circuits. In this subsection, we first introduce the class of models of our concern, which describe unitarty evolution under random Hamiltonians accompanied by randomly occurring measurements. Then we map the calculation of random-averaged purity into an equivalent quantum spin problem. In all-to-all coupled models, in which all qudits evolve and are measured under exactly the same setting (in the averaged sense), the spin model further breaks down into sectors according to permutation symmetry among qudits, which simplifies our analysis. We finally arrive at a large spin Hamiltonian, whose Hilbert space dimension is proportional to the number of qudits $N$, rather than the exponential $2^N$. 

Our starting point is a model of $N$ qudits, each with Hilbert space dimension $d$. The qudits evolve under a time-dependent Hamiltonian whose coefficients undergo Brownian motion. Specifically,
\begin{equation}    \label{eq:Brownian_ham}
    H(t)=\sum_{1\leq i_1<\dots<i_q \leq N}\sum_{a_1, \dots, a_q} J_{i_1 \dots i_q}^{a_1 \dots a_q}(t)T_{i_1 a_1} \cdots T_{i_q a_q} ,
\end{equation}
\begin{equation}
    \overline{J_{i_1 \dots i_q}^{a_1 \dots a_q}(t)}=0   ,\quad
    \overline{J_{i_1 \dots i_q}^{a_1 \dots a_q}(t)J_{i_1' \dots i_q'}^{a_1' \dots a_q'}(t')}
    =
    \mathcal{J}\delta(t-t') \delta_{i_1i_1'}\delta_{a_{1}a_{1}'} \cdots \delta_{i_qi_q'}\delta_{a_qa_q'} ,
    \label{eq:disorder are gaussian}
    \end{equation}
where $J_{i\dots}^{a\dots}(t)$'s are real, $N$ is the number of qudits, $i=1,2,...,N$ are indices for qudits, $a$'s are indices for operator basis. The expression we have written is a general $q$-body interaction. The coefficients $J_{i\dots}^{a\dots}(t)$'s undergo standard Brownian random process multiplied by strength $\mathcal{J}$. We introduced a complete orthonormal basis $\{ T_a \}$ of Hermitian operators on each qudit, satisfying $\tr T_a T_{a'} = d\delta_{aa'}$. So $a$ ranges from $0$ to $d^2-1$. For example, as to the simple case of qubits ($d=2$), we can taje these operators to be the three Pauli operators. The constant $\mathcal{J}$ labels the magnitude of the evolution, which should have reasonable scaling when $N$ and $d$ are large. The exact scaling will be discussed in later analysis. We note that all the $J_{i\dots}^{a\dots}(t)$'s are uncorrelated unless all indices coincide up to permutation. 

In addition, measurements are applied together with the unitary evolution. In each short time interval $\Delta t$, for each set of $q'$ qudits, there is probability $p = \lambda \Delta t$ that the set is subject to projective measurement $|\varphi\rangle\langle\varphi|$, where $|\varphi\rangle$ is a Haar random state drawn from the product space of the chosen qudits. $\lambda$ is a constant whose scaling dependence on $N$ and $d$ will be discussed later. As a side remark, our measurement here differs from previous works with similar setting. In Ref.~\cite{MIPT30}, a measurement of the form $\ket{\psi}\bra{\varphi}$ is considered, where $\ket{\psi}$ and $\bra{\varphi}$ are drawn from independent Haar ensembles. This can be understood as writing $|\psi\rangle\langle\varphi|=U|\varphi\rangle\langle\varphi|$ (with $U$ a Haar random matrix), which is a projective measurement followed by a random unitary operator. However, we shall see in Appendix \ref{Appendix:one cluster lindblad} that these two kinds of measurements have identical effect on the averaged purity. 

\begin{figure}[t]
    \centering
    \includegraphics[width=1\textwidth]{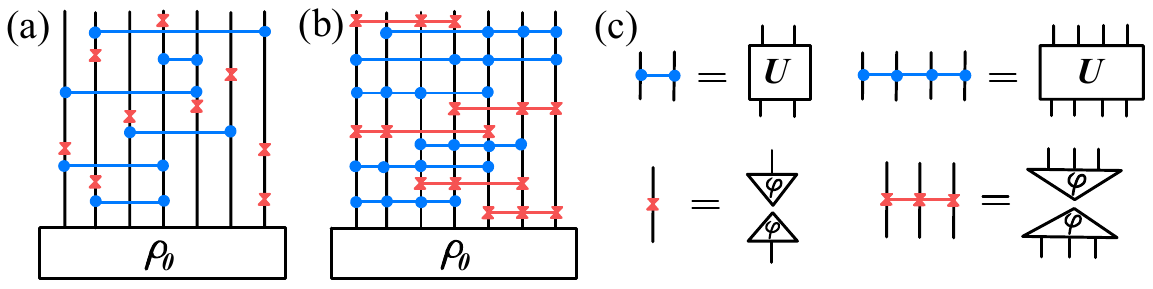}
    \caption{\label{fig:quantum circuit illustration} Illustration of examples of hybrid random circuits. (a) Two-body Brownian unitary evolution term and on-site random measurement. (b) Generic $q$-body interaction in the Brownian Hamiltonian, with $q'$-body measurement. In this figure, we take $q=4$ and $q'=3$. (c) Legends for (a) and (b). }
\end{figure}

For this family of model, we will study the second Renyi entropy as an entanglement measure. %We proceed with the entanglement measure, that is the second Rényi entropy averaged over the random evolution and measurement realizations. 
The second Rényi entropy is related to the density matrix via
\begin{equation}
    S_{A}^{(2)}(t) = - \overline{ \log \tr \rho_A^2(t) }    ,
\end{equation}
where $\rho_A(t) = \tr_{\overline{A}} \rho(t)$ is the reduced density matrix on a subsystem $A$ at time $t$. Let $V$ represent the joint evolution operator of both the unitary part and the measurement, so the unnormalized density matrix reads $\sigma(t) = V \rho_0 V^\dagger$. Noting that the projective measurement does not preserve trace, so that the density matrix needs to be renormalized: $\rho(t) = \sigma(t)/\tr \sigma(t)$. If the norm ${\rm tr}\sigma(t)$ and the second moment ${\rm tr}\sigma_A^2(t)$ both have small fluctuation in the large $N$ limit, we can approximate\footnote{\label{footnote:2}This approximation is valid when the averaging over purity ($-\log\overline{\tr\rho_A^2}$) and averaging entropy ($\overline{-\log\tr\rho_A^2}$) are close. When the fluctuation is large, $\overline{P_A}$ will favor large purity (small entropy) realizations but $\overline{-\log P_A}$ will favor large entropy realizations. We do not have a precise statement that under which condition these two averages disagree, but we will show an explicit example in Section~\ref{ssec:u1} where they disagree. }

%Furthermore, it has been believed and justified that we may move the average to be on the nominator and the denominator separately. The difference between this modified expression and the original averaged entropy is suppressed when the system size is large. Hence we may write
\begin{equation}    \label{eq:entropy_normalization}
    e^{-S^{(2)}_A(t)}
    \simeq  \overline{\tr \rho_A^2(t)}
    = \overline{\left(\frac{\tr \sigma_A^2(t)}{[\tr \sigma(t)]^2}\right)}
    \approx \frac{\overline{\tr\sigma_A^2(t)}}{\overline{[\tr\sigma(t)]^2}}
    \equiv \frac{P_A(t)}{P_{\varnothing}(t)}    .
\end{equation}
Hereafter, we focus on the unnormalized purity $P_A(t) = \overline{ \tr \sigma_A^2(t) }$, since it depends linearly on the initial state $\rho_0\otimes\rho_0$. Since the projective measurement does not preserve trace, the normalization factor $P_\varnothing (t)$ (by setting the subsystem $A$ to be the empty set $\varnothing$) must be explicitly kept. Next, we use the Choi-Jamiolkowski isomorphism (a correspondence between operators and vectors) to vectorize the doubled matrices, so the purity can be written as state overlap. Explicitly, we first introduce two states in four-replica space:
\begin{equation}
    \pket{I^+}
    = \sum\limits_{a,b=1}^{d} \ket{a}\otimes\ket{a}\otimes\ket{b}\otimes\ket{b}  ,\quad
    \pket{I^-}
    = \sum\limits_{a,b=1}^{d} \ket{a}\otimes\ket{b}\otimes\ket{b}\otimes\ket{a}  .
\end{equation}
The four replicas will later be labeled by $1$, $\bar{1}$, $2$, $\bar{2}$ in order, respectively. $\sigma\otimes\sigma$ is mapped to the state
\begin{equation}
    \sigma\otimes\sigma \rightarrow 
    \pket{\sigma\otimes\sigma} = \left( \sigma\otimes \mathbbm{1} \otimes \sigma \otimes \mathbbm{1} \right) \bigotimes_{i=1}^{N} \pket{I_i^+}   .
\end{equation}
Now the purity can be written as
\begin{equation}
    % \begin{aligned}
    P_A(t)
    = \overline{\tr[\sigma_A(t)^2]}
    = \tr[X_A\overline{\sigma(t)\otimes\sigma(t)}]\equiv(X_A|\overline{\sigma(t)\otimes\sigma(t)})  ,
    % &=(X_A|\overline{V\otimes V^*\otimes V\otimes V^*}|\rho_0\otimes\rho_0)
    \label{eq:definition of purity}
    % \end{aligned}
\end{equation}
where $X_A$ is the SWAP operator, which exchanges replica $1$ and $2$ for sites in $A$ and does nothing on $\bar{A}$ (the complement of $A$). The state equivalent to $X_A$ is 
\begin{equation}
    \pbra{X_A} = \left[\bigotimes\limits_{i\in A} \pbra{I_i^-} \right] \cdot \left[\bigotimes\limits_{i\notin A} \pbra{I_i^+} \right]  ,
\end{equation}

In principle, we are going to deal with an ensemble of random realizations of the prescribed model. But we will see that after the average, the evolution of purity simplifies into a nice form: the purity depends \emph{only} on the purity itself at an infinitesimal time before. This is a consequence of the fact that the time evolution as an ensemble is invariant under local unitary transformations on each site.

Thanks to the Markovian nature of both the evolution and the measurement, we expect a differential equation that governs all purity evolution:
\begin{equation}
    \frac{dP_A(t)}{dt} = \sum_B (\tilde{\lind})_{AB} P_B(t) ,
\end{equation}
where $A$ and $B$ are subsystems, and the sum is over all possible choices of subsystems. 

For example, we write a four-contour averaged evolution as ($\Delta t$ is infinitesimal)
\begin{equation}
    \begin{aligned}
        \overline{\pket{\sigma(t+\Delta t)^{\otimes 2}}}
        &= \overline{ e^{-iH(t)\Delta t} \otimes e^{iH^*(t)\Delta t} \otimes e^{-iH(t)\Delta t} \otimes e^{iH^*(t)\Delta t} }   \\
        &\phantom{=} \cdot \left[ \mathbbm{1} + \lambda\Delta t \sum_{\mathbf{i}}\left( \overline{\ket{\varphi_{\mathbf{i}}\varphi_{\mathbf{i}}^*\varphi_{\mathbf{i}}\varphi_{\mathbf{i}}^*}\bra{\varphi_{\mathbf{i}}\varphi_{\mathbf{i}}^*\varphi_{\mathbf{i}}\varphi_{\mathbf{i}}^*}} - \mathbbm{1} \right) \right] \overline{\pket{\sigma(t)^{\otimes 2}}}   ,
    \end{aligned}
\end{equation}
where $\sum_{\mathbf{i}}$ is a shorthand for $\sum_{1\leq i_i < \dots < i_{q'} \leq N}$, standing for $q'$-body measurement, and $\varphi_{\mathbf{i}}$ is a random state on these sites. In the exponential, time-ordered product is implicitly meant. The evolution part and the measurement involve independent random variables, so we can treat them separately. Finally, we arrive at a Lindbladian superoperator that governs the replicated evolution, as
\begin{align}
    \mathbbm{1} - \mathscr{L}_U\Delta t &= \overline{ e^{-iH(t)\Delta t} \otimes e^{iH^*(t)\Delta t} \otimes e^{-iH(t)\Delta t} \otimes e^{iH^*(t)\Delta t} }    ,\\
    \mathbbm{1} - \mathscr{L}_M\Delta t &= \mathbbm{1} + \lambda\Delta t \sum_{\mathbf{i}}\left( \overline{\ket{\varphi_{\mathbf{i}}\varphi_{\mathbf{i}}^*\varphi_{\mathbf{i}}\varphi_{\mathbf{i}}^*}\bra{\varphi_{\mathbf{i}}\varphi_{\mathbf{i}}^*\varphi_{\mathbf{i}}\varphi_{\mathbf{i}}^*}} - \mathbbm{1} \right)    ,\\
    \overline{\pket{\sigma(t)^{\otimes 2}}} &= e^{-\mathscr{L} t} \overline{\pket{\sigma(0)^{\otimes 2}}} ,\,
    \mathscr{L} = \mathscr{L}_U + \mathscr{L}_M   .
\end{align}
A detailed derivation of the explicit matrix form of this Lindbladian can be found in Appendix \ref{Appendix:one cluster lindblad}. An important observat/ion from the detailed derivation is that, although $\mathscr{L}$ is an operator in the fourfold replicated space, it induces an endomorphism on the subspace spanned by $\{ \pket{X_A} \}$. That is, we have a matrix representation of $\mathscr{L}$:
\begin{equation}    \label{eq:matrix_rep}
    \pbra{X_A} \mathscr{L} = \sum_B \pbra{X_B} (\tilde{\lind})_{AB}   ,
\end{equation}
so that
\begin{equation}
    \frac{dP_A(t)}{dt}
    = \pbra{X_A} \frac{d}{dt} \overline{\pket{\sigma(t)^{\otimes 2}}}
    = \pbra{X_A} (-\mathscr{L}) \overline{\pket{\sigma(t)^{\otimes 2}}}
    = - \sum_B (\tilde{\lind})_{AB} P_B(t)\label{eq: Lindbladian_PA}
\end{equation}
Here, we use $(\tilde{\lind})_{AB}$ for the matrix representation to reserve the symbol $(\lind)_{AB}$ for later use. On each site, we interpret the two states $\pket{I^{\pm}_i}$ as spin up and down basis. Specifically, suppose we are to calculate the purity of a subsystem $A$, then spin up at $i$ means site $i$ is not in $A$; spin down at $i$ means site $i$ is in $A$. In this representation, the Lindbladian-like equation is equivalent to the imaginary-time evolution of $N$ spin-half qubits. 

In addition to the local unitary invariance, the model we consider also has permutation invariance between different qudits, also in the sense of ensemble average. Consequently, $P_A$ and $(\tilde{\lind})_{AB}$ in Eq. (\ref{eq: Lindbladian_PA}) only depends on the size of regions $A,B$, which takes the value of $0,1,2,...,N$. In the language of spin half qubits, this means we can decompose the $2^N$ dimensional Hilbert space into representations of the total spin $S^{x,y,z} = \sum_i S^{x,y,z}_i$. The permutation invariant ``state" $P_A$ stays in the subspace with highest spin $(S^x)^2+(S^y)^2+(S^z)^2=S(S+1)$, $S=N/2$. The Lindbladian operator preserves the total spin, so that the time evolution remains in the much smaller Hilbert space of fully symmetric states, with Hilbert space dimension $N+1$. In this subspace, the $S^z$ eigenvalues corresponds to the size of subsystem $n$ in the following way:
\begin{equation}
    S^z = -\frac{N}{2}, -\frac{N}{2}+1, \dots, \frac{N}{2}
    \quad\Longleftrightarrow\quad
    n = S^z+\frac N2=0, 1, \dots, N    .
    \label{eq:dictionary}
\end{equation}
In large-$N$ limit, the leading-order terms in $N$ can be expressed in the following form (see Appendix \ref{Appendix:one cluster lindblad} for more detailed derivation):
\begin{equation}    \label{eq:lindU}
    \begin{aligned}
        \tilde{\lind}_U
        = \frac{2\mathcal{J} d^q}{q!} \Big[ d^qN^q + (2S^x)^q&-\left(\frac{d}{2}N+S^x-iS^y+dS^z\right)^q\\
        &-\left(\frac{d}{2}N+S^x+iS^y-dS^z\right)^q \Big] + O(\mathcal{J} N^{q-1})  ,
    \end{aligned}
\end{equation}
\begin{equation}    \label{eq:lindM}
    \begin{aligned}
        \tilde{\lind}_M &= \frac{\lambda}{q'! (d^{q'}+1)^2} \Big[ (d^{q'}+1)^2 N^{q'}\\
        &\phantom{=}-\left( \frac{N}{2} + S^z + \frac{1}{d}(S^x-iS^y) \right)^{q'}
        -\left( \frac{N}{2} - S^z + \frac{1}{d}(S^x+iS^y) \right)^{q'}\\ 
        &\phantom{=}-\left( S^x-iS^y + \frac{1}{d}(\frac{N}{2}+S^z) \right)^{q'}
        -\left( S^x+iS^y + \frac{1}{d}(\frac{N}{2}-S^z) \right)^{q'} \Big] + O(\lambda N^{q'-1})  ,
    \end{aligned}
\end{equation}
We remark that the constant terms in the Lindbladian can be dropped, because it cancels with itself in the denominator of Eq.~\eqref{eq:entropy_normalization}, $P_{\varnothing}(t)$. 

%We have seen that the purity evolution is equivalent to the imaginary-time evolution of a spin-half many-body system. Despite similar correspondence between random circuit models and statistical models established in previous works, our model enables further simplification on the spin model side. Thanks to the permutation-symmetric structure, we are able to decompose the Lindbladian by spin irreducible representations, eventually turning it into a classical spin problem when $N\to\infty$. Specifically, we may divide the space of fictitious spins, which is $2^N$-dimensional, according to the Casimir operator $S(S+1) = (S^x)^2+(S^y)^2+(S^z)^2$, $S$ taking values in integers or half integers. If we start in a specific sector of $S$, the time evolution never escapes this sector. Of most interest is the sector with $S=N/2$, i.e., the totally symmetric sector in which the value of purity $P_A$ only depends on the size of $A$. That is, we may introduce $P_n$ to be equal to $P_A$ with any $|A|=n$. Here, we have reduced the problem from $2^N$ dimensions to merely $(N+1)$ dimensions. A dictionary between $S^z$ eigenvalues and $n$ is
%\begin{equation}
%    S^z = \frac{N}{2}, \frac{N}{2}-1, \dots, -\frac{N}{2}
%    \quad\Longleftrightarrow\quad
%    n = 0, 1, \dots, N    .
%    \label{eq:dictionary}
%\end{equation}
However, when implementing the Lindbladian evolution of $P_n$'s, we must be careful of the normalization factor. A properly normalized eigenbasis should be $\binom{N}{n}^{-1/2} \sum_{A= |n|} P_A = \sqrt{\binom{N}{n}} P_n$, where $\binom{N}{n}$ is the binomial coefficient. Hence the equation governing $P_n(t)$ evolution is
\begin{equation}
    P_n(t) = \frac{1}{\sqrt{\binom{N}{n}}} \sum_{m=0}^N (e^{-\tilde{\lind} t})_{nm} \sqrt{\binom{N}{m}} P_m(0)    .
\end{equation}
We introduce a matrix of normalization factors $(\mathcal{N})_{mn} = \sqrt{\binom{N}{n}}\delta_{mn}$, and a column vector of purities $P = (P_0 , P_{1} , \dots , P_{N} )^T$. In matrix form, the previous equation reads
\begin{equation}    \label{eq:pur_vec_evo_nh}
    P(t) = \mathcal{N}^{-1} e^{-\tilde{\lind} t} \mathcal{N} P(0)   .
\end{equation}

In the large-$N$ limit which we are most interested in, the large spin degree of freedom becomes asymptotically classical. We are hence able to replace the operators $S^{x,y,z}$ by classical rotor on the unit sphere: $S^x\to Nx/2$, $S^y \to Ny/2$, $S^z \to Nz/2$ subject to $x^2+y^2+z^2=1$. This simplification will be implemented in detail after we Hermitize the Lindbladian in the next subsection.

\subsection{\label{ssec:pseudo-h}Pseudo-Hermiticity of the Lindbladian}

We have obtained the matrix representation regarding random unitaries and measurements, respectively, and have seen that they appear to be non-Hermitian matrices (cf. Eqs.~\eqref{eq:lindU} and \eqref{eq:lindM}). On contrary, the operators $\mathscr{L}_U$ and $\mathscr{L}_M$ are \emph{per se} self-adjoint. 

The reason of this apparent contradiction is that the basis we used to express $\tilde{\lind}$ is not orthnormal. %that we did not use an orthogonal basis in the matrix representation. 
Actually, $( I_i^{\pm} | I_i^{\pm} ) = d^2$ and $( I_i^{\pm} | I_i^{\mp} ) = d$. Consequently, $\tilde{\lind}$ is pseudo-Hermitian. A pseudo-Hermitian matrix, by definition, is an operator $A$ satisfying
\begin{equation}
    \eta A^\dagger = A \eta    ,
\end{equation}
where $\eta$ is an invertible Hermitian matrix \cite{Mostafazadeh2002pseudo,Ashida2020nonh}. Further requiring $\eta$ to be positive definite, a pseudo-Hermitian matrix can be put into Hermitian form by a similarity transformation. Consequently the entire eigenvalue spectrum of a pseudo-Hermitian matrix is real. In our case, let's define an alternative matrix for Lindbladians
\begin{equation}
    (\lind')_{AB} = \pbra{X_A} \mathscr{L} \pket{X_B}
    = \sum_C (\tilde{\lind})_{AC} ( X_C | X_B ) .
\end{equation}
We have $\pbra{X_A} \mathscr{L} \pket{X_B} = \pbra{X_B} \mathscr{L} \pket{X_A}^*$. Furthermore, we let $\eta$ be the inner products of the basis $\{ \pket{X_A} \}$. Explicitly,
\begin{equation}
    (\eta)_{AB} = (X_A|X_B),\,
    \eta = \bigotimes_{i=1}^N \eta_i  ,
\end{equation} 
where
\begin{equation}
    \eta_i
    = 
    \begin{pmatrix}
        (I_i^+|I_i^+)\  & \ (I_i^+|I_i^-)\\
        (I_i^-|I_i^+)\  & \ (I_i^-|I_i^-)
    \end{pmatrix}
    =
    \begin{pmatrix}
        d^2 & d \\ d & d^2
    \end{pmatrix}   .
\end{equation}
Alternatively, we can write $\eta_i = d \sqrt{d^2-1} e^{2\chi S_i^x}$, where $\tanh\chi = 1/d$. It follows that $\eta = (d\sqrt{d^2-1})^N e^{2\chi S^x}$ (with $S^x$ the total angular momentum). $\eta$ is Hermitian and positive definite. Now we have $\eta \tilde{\lind}^\dagger = \tilde{\lind} \eta$, and $\eta$ is positive-definite. Therefore, $\mathrm{const}\times\sqrt{\eta}$ may serve as a cure to the non-Hermitian $\tilde{\lind}$. In fact, we define
\begin{equation} \label{eq:lind_herm}
    \lind = e^{-\chi S^x} \tilde{\lind} e^{\chi S^x} = (d\sqrt{d^2-1})^{-N}\times \eta^{-1/2} \lind' \eta^{-1/2}   .
\end{equation}

The rightmost expression is evidently Hermitian. %Therefore, the ostensibly non-Hermitian Lindbladian $\tilde{\lind}$ is related to a Hermitian one $\lind$ by similarity transformation. 
This also explains why we have entirely real spectrum for $\tilde{\lind}$. The transformation $e^{\chi S^x}$ is physically a rotation around the $x$-axis by an imaginary angle. In particular, in the large-$d$ limit which we will encounter frequently, $\chi\to 0$ and therefore $e^{\chi S^x}$ becomes asymptotically trivial. The Hermitian form, Eq.~\eqref{eq:lind_herm}, makes it easier to tackle, so we shall focus on $\lind$ instead of $\tilde{\lind}$. When considering the spectrum of the Lindbladian, this is totally safe; but one should keep in mind the transformation $e^{\chi S^x}$ when recovering the purity profiles. Particularly, in the totally symmetric sector, the time evolution of the purity vector is now written as
\begin{equation}    \label{eq:pur_vec_evo}
    P(t) = \mathcal{N}^{-1} e^{\chi S^x} e^{-\lind t} e^{-\chi S^x} \mathcal{N} P(0) .
\end{equation}
This is a direct corollary of Eq.~\eqref{eq:pur_vec_evo_nh}. 

From Eqs.~\eqref{eq:lindU} and \eqref{eq:lindM}, the Hermitized form of the Lindbladians in the large-$N$ limit are
\begin{equation}    \label{eq:lindU_herm}
    \begin{aligned}
        \lind_U
        = \frac{2d^q \mathcal{J}}{q!} \Big[ (dN)^q + (2S^x)^q
        &-\left(\frac{d}{2}N+S^x+\sqrt{d^2-1}S^z\right)^q\\
        &-\left(\frac{d}{2}N+S^x-\sqrt{d^2-1}S^z\right)^q \Big]
        +O(\mathcal{J} N^{q-1}) ,
    \end{aligned}
\end{equation}
\begin{equation}    \label{eq:lindM_herm}
    \begin{aligned}
        \lind_M
        &= \frac{\lambda}{q'!(d^{q'}+1)^2} \Big[ (d^{q'}+1)^2 N^{q'} \\
        &- \left( \frac{N}{2} + \sqrt{1-d^{-2}} S^z + d^{-1} S^x \right)^{q'}
        - \left( \frac{N}{2} - \sqrt{1-d^{-2}}S^z + d^{-1}S^x \right)^{q'}  \\
        &- \left( S^x - i\sqrt{1-d^{-2}} S^y + d^{-1}\frac{N}{2} \right)^{q'}
        - \left( S^x + i\sqrt{1-d^{-2}} S^y + d^{-1}\frac{N}{2} \right)^{q'} \Big]
        + O(\lambda N^{q'-1})   .
    \end{aligned}
\end{equation}

Furthermore, if we work in the large-$d$ limit, these expressions are greatly simplified. %We shall impose a condition that each term in the Lindbladian to be proportional to $N$ in the leading order. 
Setting $\mathcal{J} = \frac{q!}{d^{2q} N^{q-1}}$ and $\lambda = \frac{q'! d^{2q'}}{ N^{q'-1} }$ and keeping the leading order ($O(N)$) terms in the Lindbladian, we finally arrive at the most concise expression of the classical model in large-$N$ and large-$d$ limit that describes the purity evolution (omitting constants):
% We rescale the constant as $\mathcal{J} \to \frac{q!\mathcal{J}}{d^{2q} N^{q-1}}$, $\lambda \to \frac{q'! d^{2q'} \lambda}{ N^{q'-1} }$, so that the Lindbladian is proportional to $N$. Finally, we arrive at the classical model in large-$N$ and large-$d$ limit that describes the purity evolution:
\begin{equation}
    \lind_U/N \approx - 2^{1-q} \cdot \left[ (1+z)^q + (1-z)^q \right] ,
\end{equation}
\begin{equation}
    \lind_M/N \approx - 2^{-q'}\cdot \left[ (1+z)^{q'}+(1-z)^{q'}+(x+iy)^{q'}+(x-iy)^{q'} \right] ,
\end{equation}
where for simplicity we have defined $a=\frac{S^a}{N/2}$ with $a=x,y,z$. When $N\to\infty$ they can be treated as components of a unit vector ${\bf n}=(x,y,z)$ satisfying $x^2+y^2+z^2=1$. 

This simple large-spin model paves the way for us to analyze the symmetries of $\lind$ and corresponding entanglement phase transitions of the steady state with analytical methods. We will present the utilization of this formulation in the next section. 

\section{\label{sec:alltoall}Entanglement phases in all-to-all models}

\subsection{\label{ssec:z2transition}$\mathbb{Z}_2$ symmetry breaking and purification transition}

Having established the quantum entanglement--classical spin mapping, we are in a place to do our first exercise to analyze the prototypical measurement-induced phase transition. We consider an $N$-qudit system under effect of simultaneous unitaries and measurements. The purity Lindbladian is taken to be
\begin{equation}    \label{eq:z2mipt_lind}
    \lind = \lind_U^{(q)} + \lind_M^{(1)} .
\end{equation}
In plain languages, there are $q$-body Brownian unitary evolution along with single-body projective measurement. Since MIPT happens only in the thermodynamic limit ($N\to\infty$), it suffices to consider only the classical rotor instead of the quantum spin. The relative strength of measurements with respect to unitaries is labeled by $\alpha$: we set $\mathcal{J} = q! d^{-2q} N^{1-q}$ in Eq.~\eqref{eq:lindU_herm} and $\lambda = d(d+1)\alpha$ in Eq.~\eqref{eq:lindM_herm}, and the full Lindbladian is the sum of these two terms. The case $q=2$ was studied in Ref.~\cite{MIPT30}. There are two apparent limits in this model. When unitaries are dominant ($\alpha \to 0$) and $q>1$, the entanglement spreads through the whole system and the final state is nearly maximally entangled, i.e., the entanglement of a subsystem of size $n$ is close to its maximal value $\min \{n, N-n\}\cdot \log d$. When measurements outnumber the unitaries ($\alpha \to \infty$), a qudit once measured cannot ever entangle again with the rest of the system, so that the entropy is greatly suppressed. The MIPT is the transition between these two phases at some finite critical measurement rate $\alpha_c$. Although both phases have volume law entropy $S\propto N$, the dependence of $S/N$ as a function of subsystem size $n/N$ changes its behavior at $n/N=1/2$. The highly entangled phase with small $\alpha$ has a derivative discontinuity of entropy $S/N$ at $n/N=1/2$, while the high $\alpha$ phase has a smooth entropy curve. In this paper, we will approach a more general version of this problem and provide an alternative interpretation of the Lindbladian. 

To begin, it behooves us to be aware of the symmetries of the Lindbladian Eq.~\eqref{eq:z2mipt_lind}. Starting from a globally pure state, per our construction of the hybrid quantum circuit model, the state remains pure in the entire evolution. This fact is obvious for the Brownian unitary part, and also not obscure with the measurements. Generally, performing a measurement on qudits $\mathbf{i}= \{ i_1, \dots, i_{q'} \}$, a state is affected like
\begin{equation}
    \ket{\psi} \quad \longrightarrow \quad \left( \ket{\varphi_\mathbf{i}} \bra{\varphi_\mathbf{i}} \otimes \mathbbm{1}_{\mathbf{i}^c} \right) \ket{\psi}
\end{equation}
where $\ket{\varphi_{\mathbf{i}}}$ denotes a Haar random state, as before, and $\mathbf{i}^c$ is the complement of $\mathbf{i}$. Hence we see a pure state is still pure after a projective measurement. This fact is addressed as the subsystem inversion symmetry, isomorphic to $\mathbb{Z}_2$. It should be noted that the total possibility decreases due to projective measurements, which is exactly why we need to divide by $P_{\varnothing}(t) = P_{n=0}(t)$ in Eq.~\ref{eq:entropy_normalization}, the defining formula for the averaged second Rényi entropy. 

Translated into spin language, the subsystem inversion symmetry means $P_n(t)=P_{N-n}(t)$ or equivalently $e^{i\pi S^{x}}P(t)=P(t)$. The Lindbladian thus obeys the following identity:
\begin{equation} \label{eq:inv_symmetry}
    e^{-i\pi S^{x}} \lind(S^x,S^y,S^z) e^{i\pi S^{x}} = \lind(S^x,-S^y,-S^z) = \lind(S^x,S^y,S^z)   .
\end{equation}
In addition, there is another independent $\mathbb{Z}_2$ symmetry which makes the full symmetry $\mathbb{Z}_2 \times \mathbb{Z}_2$. Noting that the matrix $(\lind)_{AB}$ is both Hermitian and element-wise real, the spin operator $S^y$ must appear in the form of $i S^y$ and appear in complex-conjugate pairs. Therefore, $\lind(S^x,-S^y,S^z) = \lind(S^x,S^y,S^z)$. 

Generally, when a state evolves infinite imaginary time to a final steady state, it becomes the ground state of $\lind$. We can write the following expression for the steady-state purity when the ground state is non-degenerate:
\begin{equation}
    P_n(t\to\infty) \propto \bra{n} \mathcal{N}^{-1} e^{\chi S^x} \ket{\Psi_0} ,
\label{eq:3.4}
\end{equation}
where $\ket{\Psi_0}$ is the ground state of $\lind$. Otherwise, if degenerate ground states are present, we must take into account all their contributions on the right-hand side. We conclude that the entanglement phases are directly related to the ground-state manifold of the mapped Lindbladian. 

After all the above preparations, we are now at a place to infer the entanglement phases, which correspond to different phases of ground states of $\lind$. It is straightforward to see $S^y$ inversion symmetry is never broken, i.e., $\bra{\Psi_0} S^y \ket{\Psi_0} = 0$ for ground state(s). This is because $|\Psi_0\rangle$ is a real vector in $S^z$ basis, and $S^y$ is hermitian but purely imaginary. So it's expectation value is zero. Hence the crux of matter is the spontaneous breaking of the subsystem inversion symmetry. In large $N$ limit, as we discussed earlier, we can treat $(x,y,z)=(S^x,S^y,S^z)/(N/2)$ as a classical normal vector, so that we can define the spherical coordinate $x=\cos\varphi\sin\theta$, $y = \cos\theta$, $z = \sin\varphi\cos\theta$. We just need to minimize the classical function $\lind(x,y,z) = \lind(\theta, \varphi)$. To search for the ground states, we can fix $\theta=\pi/2$ by $\langle\Psi_0|S^y|\Psi_0\rangle=0$, and focus on the sign of the second-order derivative at $\varphi=0$, which results in the critical measurement strength:
\begin{equation}    \label{eq:mipt_alphac}
    \alpha_c = 2q\cdot[d^{-q}+2^{1-q}(1+d^{-1})^{q-1}(q-1-qd^{-1})] .
\end{equation}
When $\alpha < \alpha_c$, the ground state is doubly degenerate and the whole system is in the scrambled phase; conversely, $\alpha<\alpha_c$ indicates the purified phase that has a unique ground state. Here comes the advantage of our classical-spin formulation compared to the original approach in Ref.~\cite{MIPT30}. In Ref.~\cite{MIPT30}, the entanglement dynamics is mapped to a Hamilton-Jacobi equation of a quantum particle in an inhomogeneous potential. However, its solvability strongly relies on the tridiagonal structure of the Lindbladian, thus is applicable only when $q=2$. Our method prevails both in generality and analyticity. 

In general cases, in the symmetry-breaking phase, the final state depends on the initial state. That is to say, a finite amount of initial information survives up to infinite time. This is the case if we start from an arbitrary initial state. However, if we start from a pure state whose purity observes subsystem inversion symmetry, the overlap between the initial state and the two symmetry-breaking ground states must be the same. Consequently, the final state is fixed to be a Page curve, exhibiting piecewise linearity and a cusp at half system size (see Sec.~\ref{ssec:page} for justification of these features).

\subsection{\label{ssec:page}Steady state Page curve}

To further demonstrate the power of the quantum-classical mapping, we calculate the entanglement entropy of the final state in specific cases. With the spin representation, we are able to derive explicit formulas for the entanglement entropy in numerous cases. 

\begin{figure}[h]
\centering
\includegraphics[width=1\textwidth]{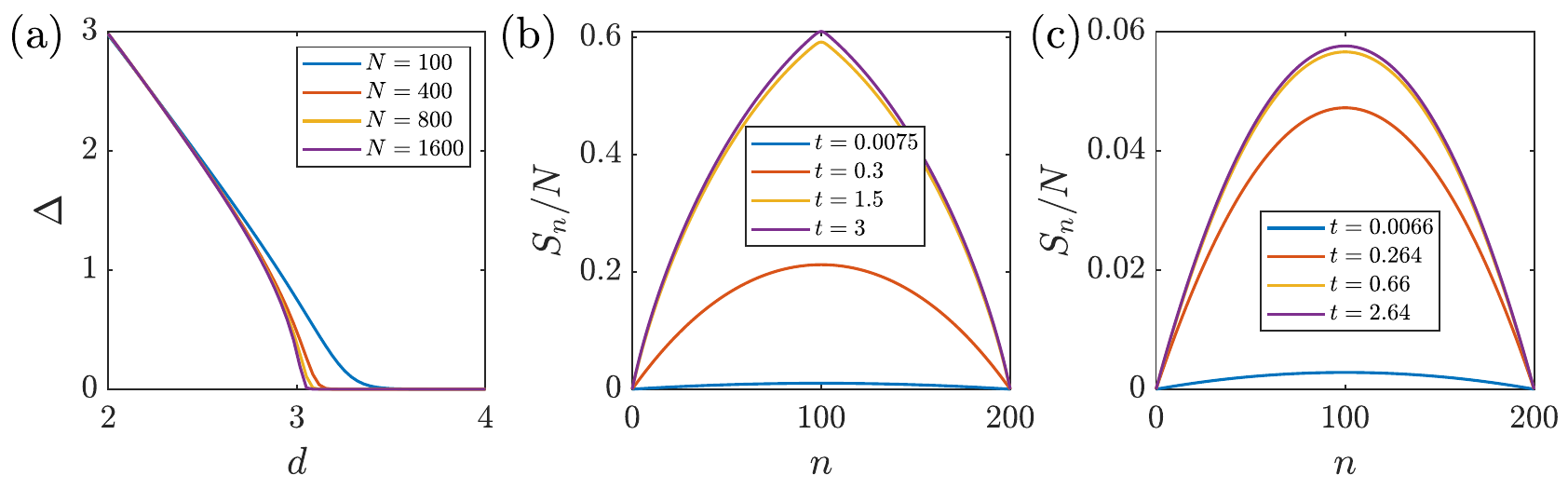}
\caption{\label{fig:MIS: gap and dynamics} Dynamics measurement-only models with three-body measurement ($q'=3$). (a) The Liouvillian gap of $\lind_M^{(3)}$ in Eq.~\eqref{eq:lindM_herm}. When system size $N$ approaches infinity, we can infer a phase transition at $d_c = 3$. (b) (c) Demonstration of real-time evolution of the second Rényi entropy, plotted as a function of subsystem size $n$. The total cluster size $N=200$. (b) is in the scrambled phase with $d=20$, while (c) is in the purified phase with $d=2$. They are distinguished visually by the appearance of a cusp. }
\end{figure}

We consider the same setting as in the previous section. We begin by showing that  for only onsite measurement ($q'=1$), the steady state have vanishing entropy. In fact, the unique ground state is always $\ket{\rightarrow}$ (the highest-weight state of $S^x$ satisfying $S^x\ket{\rightarrow} = N/2 \ket{\rightarrow}$), since the Lindbladian is the single-body measurement term $\lind_M^{(1)} \propto -x$. Hence the purity is
\begin{equation}
    \begin{aligned}
        P_n(\infty) &\propto \bra{n} \mathcal{N}^{-1} e^{\chi S^x} \ket{\rightarrow} \\
        &= \frac{e^{\chi N/2}}{\sqrt{\binom{N}{n}}} \bracket{n}{\rightarrow}  \\
        % &= \frac{e^{\chi N/2}}{\sqrt{\binom{N}{n}}} \sqrt{\frac{\binom{N}{n}}{2^N}}
        &= e^{(\chi-\log 2)N/2} ,
    \end{aligned}
\end{equation}
which does not depend on $n$. We conclude that the entanglement entropy is identically zero in the purified phase for one-body measurement-only circuit. One should be careful that $\ket{n}$ is equal to the spin $S^z$ basis vector $\ket{S^z=N/2-n}$ per our definition.  For more general parameter choices, as long as we are in purified phase, the entropy curve is $O(N)$ and $S^{(2)}_n/N$ is a smooth function of $x:=\frac{n}{N}\in[0,1]$, without cusp. This is supported by numerical simulations, c.f. figure~\ref{fig:MIS: gap and dynamics}.  

We move on to consider the entanglement entropy in the scrambled phase. We take $q=2$ as an elementary example. Let $\mathcal{J} = q! d^{-2q} N^{1-q} = 2/(d^4 N)$ and $\lambda = d(d+1)\alpha$. In this convention, the full Lindbladian reads
\begin{equation}    \label{eq:lind_original_mipt}
    \lind(x,y,z) / N = \left( \frac{x}{d} \right)^2 - (1-\frac{1}{d^2})z^2 - \frac{2x}{d} -\alpha x .
\end{equation}

First, we inspect the entropy at $\alpha=0$, i.e., no measurements but only unitaries governs the dynamics. Parametrizing $x=\cos\varphi$, $y=0$, $z=\sin\varphi$, we find two minima at $\pm\varphi_0$, where $\varphi_0 = \arccos d^{-1}$. These two directions correspond to the spin coherent states in the quantum spin problem. To be concrete, a minimum in direction $\hat{\Omega} = (x,y,z)$ corresponds to the spin coherent state $\ket{\hat{\Omega}}$. In this case, the two ground states can be expressed as
\begin{equation}
    e^{-i(\pi/2-\varphi_0)S^y}  \ket{\uparrow}
    \quad\textrm{and}\quad
    e^{i(\pi/2-\varphi_0)S^y} \ket{\downarrow}    ,
\end{equation}
where $\ket{\uparrow}$ and $\ket{\downarrow}$ are highest and lowest weight state of spin $S^z$, respectively. Furthermore, we assume that the initial state is a global pure state so that it has equal overlap with the two degenerate ground states. Working in the leading order of $d^{-1}$, we calculate the steady state purity as\footnote{To be precise, we are sending $N$ to infinity and then $d$ to infinity. Eventually we require $1\ll d \ll N$. }
\begin{equation}
\begin{aligned}
P_n(\infty)
&\propto \bra{n} \mathcal{N}^{-1}e^{\chi S^x}(e^{-i(\pi/2-\varphi_0)S^y} \ket{\uparrow}+e^{i(\pi/2-\varphi_0)S^y} \ket{\downarrow})\\
&\approx \frac{1}{\sqrt{\binom{N}{n}}} \bra{n} e^{d^{-1}S^x}(e^{-id^{-1}S^y} \ket{\uparrow}+e^{id^{-1}S^y} \ket{\downarrow})\\
&\approx \frac{1}{\sqrt{\binom{N}{n}}} \bra{n} \left[ \sum_{k=0}^{n}\frac{(S^{-}/2d)^{n-k}}{(n-k)!}\frac{(S^{-}/2d)^{k}}{k!}  \ket{\uparrow} + \sum_k\frac{(S^{+}/2d)^{n-k}}{(n-k)!}\frac{(S^{+}/2d)^{k}}{k!} \ket{\downarrow} \right] \\
&= (d^{-1})^{n}+(d^{-1})^{N-n}  .
\end{aligned}
\end{equation}
In the second line, we approximated $\chi = \operatorname{arctanh} d^{-1} \to d^{-1}$ and $\pi/2-\varphi_0 \to d^{-1}$. In the third line, we Taylor-expanded $e^{d^{-1} S^x}$ and $e^{\pm i d^{-1} S^y}$ and collected the first non-zero order in $d^{-1}$. Therefore, the entanglement entropy is 
\begin{equation}    \label{eq:page_purity_a0}
    e^{-S_n(\infty)} = \frac{P_n(\infty)}{P_0(\infty)} \approx \frac{(d^{-1})^{n}+(d^{-1})^{N-n}}{1+(d^{-1})^{N}} .
\end{equation}
%This formula shows clear characteristic of a Page curve. 
This result is identical to the entropy behavior of a random state. Assuming $N$ is very large, we have $S_n = n\log d$ when $n<N/2$, and $S_n = (N-n)\log d$ when $n>N/2$, exhibiting a cusp at $n=N/2$. This formula almost saturates the maximal possible entanglement entropy. 

Then we consider the general case of $\alpha\neq 0$. In principle, the entropy can be explicitly carried out via the spin coherent state techniques, but the result is involved due to the two-fold rotation $e^{-\chi S^x} e^{\pm i \varphi_0 S^y}$. For simplicity, we present the calculation in the large-$d$ limit, when the imaginary rotation $e^{\chi S^x}$ becomes trivial.  Eq.~\eqref{eq:mipt_alphac} for $q=2$, $d\to\infty$ leads to a phase transition at $\alpha_c = 2$. In the scrambled phase $\alpha<2$, the minimum of $\lind$ is reached at $\varphi = \pm\varphi_0$, where $\varphi_0=\arccos (\alpha/2)$; while in the purified phase $\alpha>2$, there is a unique minimum at $\varphi=0$. Similar to the previous calculation, when $\alpha<2$ we carry out the calculation of purity as
\begin{equation}
\begin{aligned}
P_n(\infty)
&\approx \bra{n} \mathcal{N}^{-1}\left(e^{-i(\pi/2-\varphi_0) S^y} \ket{\uparrow}+e^{i (\pi/2-\varphi_0) S^y} \ket{\downarrow}\right)    \\
&= \frac{1}{\sqrt{\binom{N}{n}}} \cdot \sqrt{\frac{N!}{(N-n)!n!}}\cdot\bigg[\left(\cos{\left(\frac{\pi}{4}-\frac{\varphi_0}{2}\right)}\right)^{N-n}\left(\sin{\left(\frac{\pi}{4}-\frac{\varphi_0}{2}\right)}\right)^{n} \\
&\phantom{=} + \left(\cos{\left(\frac{\pi}{4}-\frac{\varphi_0}{2}\right)}\right)^{n}\left(\sin{\left(\frac{\pi}{4}-\frac{\varphi_0}{2}\right)}\right)^{N-n}\bigg]   ,
% &= \sum_{n=0}^{N} \bigg[\left(\cos{\left(\frac{\pi}{4}-\frac{\varphi_0}{2}\right)}\right)^{N-n}\left(\sin{\left(\frac{\pi}{4}-\frac{\varphi_0}{2}\right)}\right)^{n} \\
% &\phantom{=} + \left(\cos{\left(\frac{\pi}{4}-\frac{\varphi_0}{2}\right)}\right)^{n}\left(\sin{\left(\frac{\pi}{4}-\frac{\varphi_0}{2}\right)}\right)^{N-n}\bigg] \ket{n}   ,
\end{aligned}
\label{eq:cusp_expression}
\end{equation}
where we expanded the spin coherent state using the Schwinger boson representation \cite{book2}. We finally arrive at the entanglement entropy:
\begin{equation}    \label{eq:page_purity_larged}
    e^{-S_n(\infty)}=\frac{P_n(\infty)}{P_0(\infty)}=\frac{(\tan{(\frac{\pi}{4}-\frac{\varphi_0}{2})})^{n}+(\tan{(\frac{\pi}{4}-\frac{\varphi_0}{2})})^{N-n}}{1+(\tan{(\frac{\pi}{4}-\frac{\varphi_0}{2})})^{N}}    .
\end{equation}
This formula resembles Eq.~\eqref{eq:page_purity_a0} in structure, indicating qualitatively similar Page curve entropy. The ``effective Hilbert space dimension'' is $d_{\text{eff}}=\cot\left(\frac{\pi}4-\frac{\varphi_0}2\right)$. For more general $q$, the scrambling phase always turns out to have Page-like entropy, with a modified $\varphi_0$ that depends on $q$. 

% On the other hand, the entropy in the purified phase may be inferred from exactly the same derivation. But substituting $\varphi_0=0$ in Eq.~(\ref{eq:page_purity_larged}) we get a vanishing purity, indicating that the large-$d$ approximation may be too rough. To fix this, we directly set the unique ground state to be $\ket{\rightarrow}$, which is the highest-weight state of $S^x$: $S^x\ket{\rightarrow} = N/2 \ket{\rightarrow}$. 

\subsection{\label{ssec:MIS}Measurement-induced scrambling}

Although measurement has the effect of purifying a state, measurements involving multiple qudits can also induce entanglement between these qudits~\cite{PhysRevX.11.011030}. In this section, we consider the entanglement phases in a measurement-only system. Obviously, %At the beginning, one notices the the obvious power of measurement to purify a state. A set of qudits measured turns into a local pure state, and thus disentangle from the rest part of the system. 
if all the measurements applied commute with each other, the final state is inevitably purified. This is the case in Secs.~\ref{ssec:z2transition} and \ref{ssec:page}, where single-body measurement showing only the power of purification. Conversely, when some measurements do not commute with each other, a subsystem purified by a measurement may re-entangle with the rest of the system by subsequent measurements. Indeed, in reference \cite{PhysRevX.11.011030}, the author find numerical evidence that certain level of measurement ``frustration'' is necessary to generate a globally scrambled phase using only measurement.  However, an analytical characterization of the conditions under which the measurement-only circuit would lead to scrambling instead of purification has not been fully developed. 

\begin{figure}[t]
    \centering
    \includegraphics[width=0.85\textwidth]{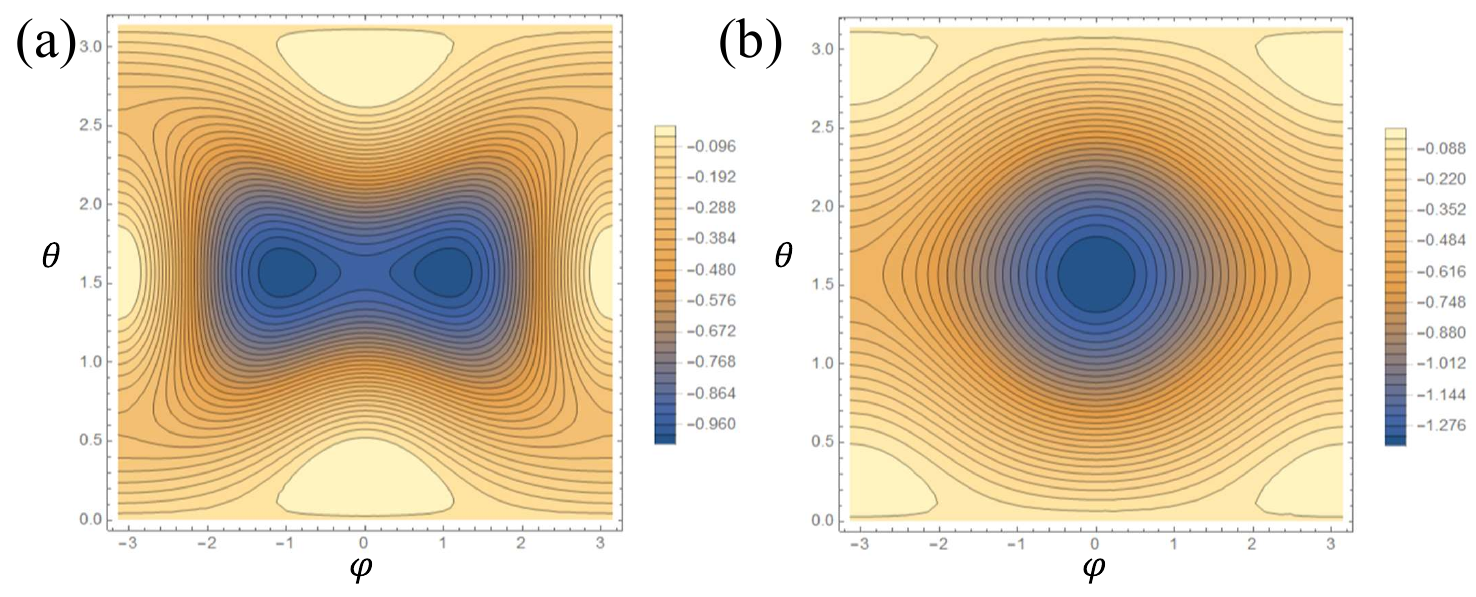}
    \includegraphics[width=0.85\textwidth]{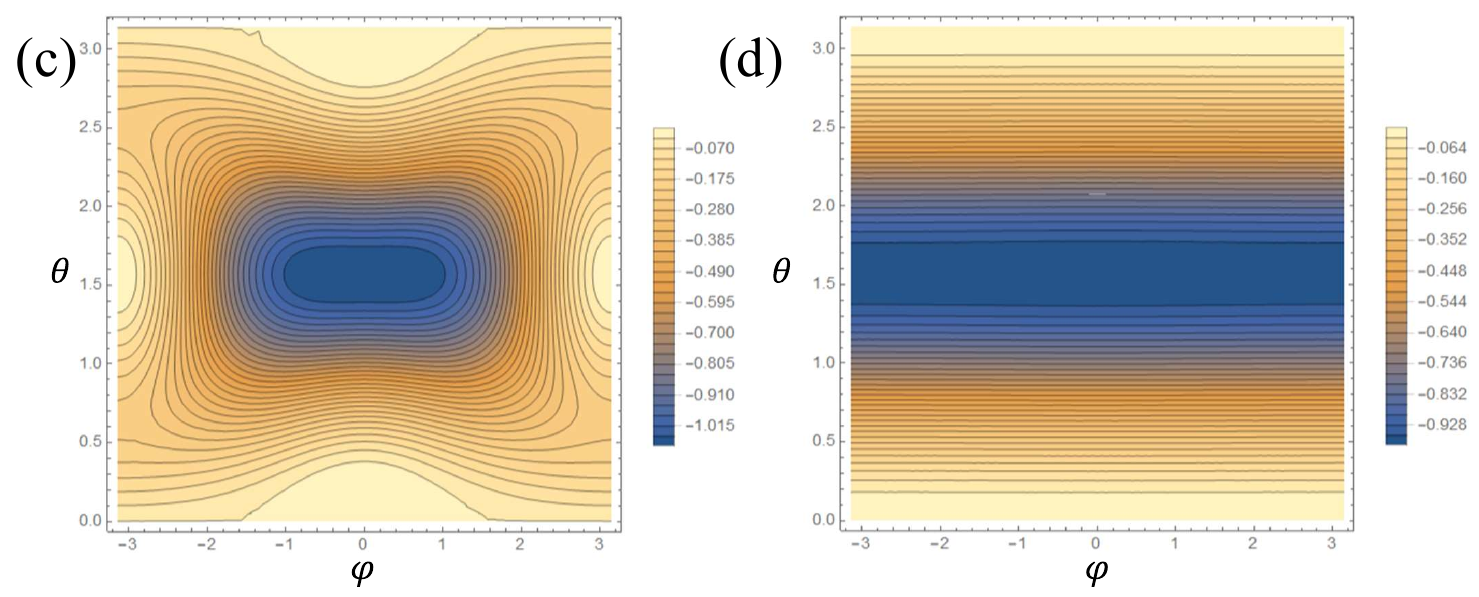}
    \caption{\label{fig:largeS landscape} Measurement-only Lindbladians $\lind_M^{(q')}$ in Eq.~\eqref{eq:lindM_herm}, plotted as a function of spherical coordinates: $\lind_M( x=\sin\theta\cos\varphi, y = \cos\theta, z=\sin\theta\sin\varphi )$. All the presented cases agree with the prediction of the critical value of $d$ in Eq.~\eqref{eq:MIS_crit_d}. (a) $q'=3$, $d=4$. degenerate ground states indicate scrambled phase. (b) $q'=3$, $d=2$. Unique ground state indicates purified phase. (c) $q'=3$, $d=d_c=3$. This is the critical point between case (a) and (b), showing a purified phase but with critical relaxation time. (d) $q'=2$, $d=d_c=\infty$. A $\mathrm{U}(1)$ symmetry emerges on the equator, also indicating a critical purification time. }
\end{figure}

We will look in detail $\lind = \lind_M^{(q')}$ with definite $q'$ (see Eq.~\eqref{eq:lindM_herm}). More generally, we may consider linear combination of terms with different $q'$'s, but it is essential to first understand the effect of them separately. Within our formulation, the phase transition is identified with the spontaneous breaking of subsystem inversion symmetry. Numerical calculation of some typical cases of $\lind_M^{(q')}$ as a function of $\theta$ and $\varphi$ is presented in Fig.~\ref{fig:largeS landscape}. Through examining the second derivative at $x=1$ (or equivalently $\varphi=0$), the phase boundary is found to be
\begin{equation}    \label{eq:MIS_crit_d}
    d_c(q') = \frac{q'}{q'-2}   .
\end{equation}
When the dimension of the qudit $d$ is below the threshold $d_c$, the system is in the purified phase and the equivalent spin model has a unique ground state; while for $d>d_c$, it is in the scrambled phase with doubly degenerate ground states. Interestingly, ``frustration'' is not sufficient for reaching a highly entangled state with only measurement. Two-body measurement, although still being non-commutative, is insufficient for scrambling phase, since $d_c=\infty$ for $q'=2$. The minimal $q'$ required for a scrambling phase decreases upon increasing $d$. Especially, at least five-body measurement is needed for quotidian qubits ($d=2$, $q'_{\min} = 5$). $d=2$, $q'=4$ lies on the phase boundary, suggesting a critical entanglement behavior. We will elaborate on the critical behaviors in Sec.~\ref{ssec:u1} and Appendix \ref{app:gap}. 

In Fig.~\ref{fig:MIS: gap and dynamics}, we provide numeric results regarding the time evolution of entanglement. We set $q'=3$ and demonstrate the gap between the two lowest eigenvalues $E_1-E_0$ in (a), the entropy evolution in the scrambled phase in (b), and the purified phase in (c). The scrambled phase is easily recognized with a cusp in the entropy profile at late time. In Fig.~\ref{fig:largeS landscape}, one can see the spontaneous symmetry breaking corresponding to Eqs.~\eqref{eq:lindM_herm} and \eqref{eq:MIS_crit_d}. 

Finally, we remark on the finite-time dynamics of the entanglement purity. The time scale needed to approach the final steady state is directly related to the so-called Liouvillian gap, i.e., the eigenvalue gap between the ground-state manifold and the first excited state of the entanglement Lindbladian $\lind$. Denoting the Liouvillian gap by $\Delta$, then any initial state will be sufficiently close to the steady state after a relaxation time $1/\Delta$. We find that the gap for $\lind_M^{(q')}$ is finite in the purified phase, but is proportional to  $N^{-1}$ at the critical parameter $d=d_c$ in Eq.~\eqref{eq:MIS_crit_d}. The latter situation is what we call the critical purification. To derive the aforementioned Liouvillian gap scaling, we need to work in the quantum picture and expand around the classical solution $(x,y,z) = (1,0,0)$. The detailed procedure can be found in Appendix \ref{app:gap}.

%XLnote: unfinished here
\subsection{\label{ssec:u1}Emergent $\mathrm{U}(1)$ for two-body measurement}

In this section we study a peculiar model that exhibits continuous $\mathrm{U}(1)$ symmetry. We set $q'=2$ and $d\to\infty$. The entanglement purity Lindbladian reads (see Eq.~\eqref{eq:lindM_herm} for the general formula)
\begin{equation}    \label{eq:lind_U1}
    \lind_M^{(2)} = \frac{1}{2} \left( -x^2-z^2+y^2 \right) + O(d^{-1}) .
\end{equation}
We observe that the $\mathrm{U}(1)$ symmetry, as a rotation in the $x$-$z$ plane, is exact in the large-$d$ limit. This symmetry is seen graphically in Fig.~\ref{fig:largeS landscape}(d), represented by the independence of $\lind_M^{(2)}$ on $\varphi$. 

The $\mathrm{U}(1)$ symmetry is responsible for the critical purification time that is proportional to $N$, since the Liouvillian gap is $O(N^{-1})$. This can be understood easily from the quantum picture. The quantum expression of Eq.~\eqref{eq:lind_U1} is
\begin{equation}
    \lind_M^{(2)} \sim N^{-1} (S^y)^2
\end{equation}
after dropping constants and $O(d^{-1})$ terms. For simplicity, we assume that $N$ is even. Then the ground state is the state annihilated by $S^y$, and the first excited states are $S^y = \pm 1$ eigenstates. It is clear now that the Liouvillian gap is $O(N^{-1})$. This is in sharp contrast with the normal purified phase which exhibits an $O(1)$ Liouvillian gap, as can be seen from the simplest Lindbladian $\lind_M^{(1)} \sim -S^x$. In Appendix \ref{app:gap}, we find for general cases that the gap scaling is $O(1)$ for  purified phase, and $O(N^{-1})$ at the phase transition point. 

The entanglement entropy of the steady state of this $\mathrm{U}(1)$ model is also interesting by itself, since it differs from all typical cases aforementioned. For simplicity, we consider the case that $N$ is even, so that the ground state is unique. The ground state satisfies $S^y\ket{\Psi_0} = 0$, so its components $a_m \equiv \bracket{m}{\Psi_0}$ can be determined recursively, using
\begin{equation}
    \frac{S^+ - S^-}{2i} \ket{\Psi_0} = 0
    \,\Longrightarrow\,
    a_{m-1}\sqrt{S(S+1)-m(m-1)}=a_{m+1}\sqrt{S(S+1)-m(m+1)} .
\end{equation}
Through the purity $P_n = \bra{n} \mathcal{N}^{-1} \ket{\Psi_0}$, the second Rényi entropy is expressed as
\begin{equation}
    S^{(2)}_{n}(t=+\infty)
    = \log\left((\mathcal{N})_{nn}\right)-\log\left(\frac{a_{S-n}}{a_S}\right)  .
\end{equation}
It is easily known that $a_{S-n} = 0$ for odd $n$, such that $S_n^{(2)} = \infty$ for odd $n$. While for even $n$, $a_{S-n}$ is finite, and the derivation proceeds as

\begin{equation}    \label{eq:U1_entropy}
    \begin{aligned}
        S^{(2)}_{n}(t=+\infty)
        &= \frac{1}{2}\log\left( \frac{N!}{n!(N-n)!} \right)+\frac{1}{2}\log \frac{\Gamma(\frac{N}{2}+1)\Gamma(1)\Gamma(\frac{N}{2}-\frac{n}{2}+\frac{1}{2})\Gamma(\frac{n}{2}+\frac{1}{2})}{\Gamma(\frac{N}{2}+\frac{1}{2})\Gamma(\frac{1}{2})\Gamma(\frac{N}{2}-\frac{n}{2}+1)\Gamma(\frac{n}{2}+1)}  \\
        &\approx -\frac{N}{2}\left[\left(\frac{n}{N}\right)\log\left(\frac{n}{N}\right) + \left(1-\frac{n}{N}\right)\log\left(1-\frac{n}{N}\right) \right] + \frac{1}{4}\log\left( \frac{n(N-n)}{2N}\right)\\
        &\approx - \frac{N}{2}[p\log p + (1-p)\log (1-p)]+O(\log N)    ,
    \end{aligned}
\end{equation}
where $p=n/N$, and $\Gamma(z)$ is the gamma function satisfying $\Gamma(z+1) = z\Gamma(z)$. In the second line, we used Stirling's approximation $\log(n!) = n\log n-n+O(\log n)$. In the third line, we noticed that the term $- \log (a_{S-n}/a_S)$ is $O(\log N)$, so is significantly smaller than the contribution of $\log\sqrt{\binom{N}{n}} = O(N)$. We conclude that when $n$ is even, the entropy curve is roughly that of a binomial distribution with probability $p=n/N$, multiplied by $N/2$. 

Finally, we provide an intuitive interpretation for the result Eq.~\eqref{eq:U1_entropy}. This discussion will serve as an example that averaging over purity and average over entropy will have significant difference (see dicussion in footnote~\ref{footnote:2}.) The entanglement is ``dimerized'' in the final state. More specifically, each two qudits in the whole cluster are likely to form a highly entangled pair (becoming maximally entangled in large $d$ limit) , due to the two-body measurements. This is reflected qualitatively in the expression of the entropy. When calculating $S_n^{(2)}$, we are bisecting the whole system into a subsystem $A$ of size $n$ its complement $\bar{A}$ of size $N-n$. For a single choice of $A$, if $A$ separate at least one entangled pair, then in large $d$ limit it has purity $P_A\approx0$, otherwise $P_A\approx1$. For even $n$, the averaged purity ($P_n:=\frac{1}{n}\sum_{A,|A|=n}P_A$) becomes a counting problem: there are $\binom{N/2}{n/2}$ ways of choosing $A$ that doesn't `break' any entangled pair. So we obtain average purity being $P_n\propto\binom{N/2}{n/2}$, which is exactly the $O(N)$ part of~\eqref{eq:U1_entropy}. For odd $n$ however, the subsystem $A$ must `break' at least one entangled pair, so we have $P_A\approx0\  \forall A$, therefore $P_n\approx0, S_n^{(2)}\approx+\infty$.

\section{\label{sec:spatial}Entanglement phases with spatial structure}

Having been acquainted to the formalism of mapping quantum entanglement dynamics to large-spin models, we are now in a place to generalize it to qudits with spatial structure. Specifically, the subject in the last section, with all-to-all interaction, can be viewed as a zero-dimensional quantum dot after mapping. Endowed with a lattice spatial arrangement, the entanglement phases are expected to be far more versatile. 

In Sec.~\ref{ssec:2interaction}, we analyze the possible ways two clusters of qudits may couple. In Sec.~\ref{ssec:scramble_hiD} we dive into concrete examples of hybrid quantum circuits with spatial locality, and then we discuss a specific measurement-only circuit in which an $\mathrm{SU}(2)$ symmetry emerges in Sec.~\ref{ssec:su2}.

\subsection{\label{ssec:2interaction}Two-cluster interaction}

\begin{table}[t]  
    \footnotesize
    \begin{center}   
    \begin{tabular}{c|c|c|c|c|c|c}   
    \hline   Type & Locality & Infinite-$d$ Lindbladian & Steady states & Degeneracy & \makecell[c]{Entanglement\\phase} & \makecell[c]{Spin language} \\   
    \hline   
    $\lind_U^{(q)}$ & $q$ & $-(1+z)^q-(1-z)^q$ & $z=\pm1$ & 2 & Scrambled & \makecell[c]{Two-fold\\crystal field}   \\ 
    \hline   
    $\lind_M^{(1)}$ & $1$ & $-x$ & $x=1$ & 1 & Purified & Zeeman field\\ 
    \hline   
    $\lind_M^{(2)}$ & $2$ & $-x^2-z^2+y^2$ & $y=0$ & $1^*$ & \makecell[c]{Critical\\purified} & $\mathrm{U}(1)$ field\\ 
    \hline   
    $\lind_M^{(q')}$ & $q'\geq3$ & \makecell[c]{$-(1+z)^{q'}-(1-z)^{q'}$\\$-(x+iy)^{q'}-(x-iy)^{q'}$} & $z=\pm1$ & 2 & Scrambled & \makecell[c]{Two-fold\\crystal field}\\ 
    \hline   
    $\lind_U^{(q,q)}$ & $(q,q)$ & \makecell[c]{$-(1+z_1)^q(1+z_2)^q$\\$-(1-z_1)^q(1-z_2)^q$} & $z_1 = z_2 = \pm 1$ & 2 & Scrambled & \makecell[c]{$\mathbb{Z}_2$-symmetric\\coupling}    \\ 
    \hline   
    $\lind_M^{(1,1)}$ & $(1,1)$ & $-x_1x_2-z_1z_2+y_1y_2$ & Spin singlet  & $1^*$ & \makecell[c]{Critical\\purified} & \makecell[c]{AFM\\Heisenberg}    \\
    \hline   
    $\lind_M^{(q',q')}$ & \makecell[c]{$(q',q')$\\$q'\geq2$} & \makecell[c]{$-(1+z_1)^q(1+z_2)^q$\\$-(1-z_1)^q(1-z_2)^q$\\$-(x_1+iy_1)^q(x_2+iy_2)^q$\\$-(x_1-iy_1)^q(x_2-iy_2)^q$} & $z_1 = z_2 = \pm 1$ & 2 & Scrambled & \makecell[c]{$\mathbb{Z}_2$-symmetric\\coupling}   \\
    \hline   
    \end{tabular}  
    \caption{Summary of the unitary or measurement Lindbladians considered in this paper. Each term involve either one or two clusters, and the superscripts of $\lind$ is the number of qudits that an operation involves. For example, $\lind_U^{(q,q)}$ means the Brownian unitary effect, such that the Hamiltonian involves $q$ qudits from cluster 1 and $q$ qudits from cluster 2. For the asterisked ($^*$) items in ``Degeneracy'', the infinite-$d$ expressions respect $\mathrm{U}(1)$ or $\mathrm{SU}(2)$ continuous symmetry, so their ground state manifold should be a continuum. But the degeneracy is lifted when $d$ is finite. 
    }
    \end{center}   
    \label{table:summary}
\end{table}

In parallel with the construction of the unitary term and the measurement term in Sec.~\ref{sec:setup}, here we will introduce their generalizations that involve two distinct clusters of qudits. Such construction is a basic ingredient in considering the effect of inter-cluster interactions in a lattice of qudit clusters. 

Illustration of our idea of two-cluster unitary and two-cluster measurement are shown in Fig.~\ref{fig:cluster illustration} (b) and (c), respectively. Consider two clusters of qudits, each having $N$ qudits of dimension $d$\footnote{We assume the two clusters has the same $N$ and $d$ for simplicity, but one can generalize our expressions to non-equivalent clusters without effort. }. A $(q, q)$-body Brownian Hamiltonian is defined as
\begin{equation}
    H(t) = \sum\limits_{\textbf{i},\textbf{j}}\sum_{\textbf{a},\textbf{b}}J_{\textbf{i},\textbf{j}}^{\textbf{a},\textbf{b}}(t)T_{i_1}^{a_1}\cdots T_{i_q}^{a_q}\cdot T_{j_1}^{b_1}\cdots T_{j_q}^{b_q}  ,
\end{equation}
\begin{equation}
    \overline{J_{\mathbf{i},\mathbf{j}}^{\textbf{a},\textbf{b}}(t)}=0,\ \overline{J_{\textbf{i},\textbf{j}}^{\textbf{a},\textbf{b}}(t)J_{\textbf{i'},\textbf{j'}}^{\textbf{a'},\textbf{b'}}(t')}
    = \mathcal{J} \delta(t-t')\delta_{\textbf{i}\textbf{i'}}\delta_{\textbf{j}\textbf{j'}}\delta_{\textbf{a}\textbf{a'}}\delta_{\textbf{b}\textbf{b'}}  ,
\end{equation}
where $\mathbf{i} \in \{(i_1,\dots,i_q)| 1\leq i_1<\dots<i_q \leq N \}$, $\mathbf{a} \in \{(a_1,\dots,a_q)|a=0,\dots,d^2-1\}$ are indices in the first cluster, and $\mathbf{j}$, $\mathbf{b}$ similarly for the second. The construction of $(q', q')$-body measurement is similar. We draw $q'$ qudits in cluster one and two each, and form a Haar random state $\ket{\varphi}$ in the space spanned by these qudits. In a short time period, there is a possibility that a projective measurement $\ket{\varphi}\bra{\varphi}$ is performed on the physical state. The Lindbladian corresponding to two-cluster unitary or measurement are readily obtained by similar reasoning as in Sec.~\ref{sec:setup}. We present their detailed derivation in Appendix \ref{Appendix:two cluster lindblad}. Our focus will be the leading terms in the large-$d$ limit, and they are listed in Table \ref{table:summary}. Each type of Lindbladian is labeled by the number of qudits in the first and the second sector that an operation involves. The reason why we only present the Lindbladian in large-$d$ limit is that they have simple expressions, and they are sufficient for reading out the ground state degeneracy, the local minima, etc. In particular, the one-one body measurement $\lind_M^{(1,1)}$ corresponds to a Heisenberg-type interaction, and has exceptionally large symmetry group. This can induce unexpected critical entanglement behavior, which we will discuss in detail in Sec.~\ref{ssec:su2}. 

As a first exercise of the above constructions, we study two coupled clusters with both unitaries and measurements, which can exhibit more abundant phases than single-cluster models. One finds that the inter-cluster interaction strength affects the purification efficiency (whether one can purify the whole system by measuring only one cluster). The phase diagram of two clusters is calculated and demonstrated in Appendix \ref{appendix:example of two cluster phase diagram}. 

This two-cluster model is interesting by itself, since it may be interpreted as a prototype of more involved entanglement phases when the clusters are endowed with lattice structure, which we will elaborate on in the following sections. 

% Two coupled cluster model can exhibit abundant entanglement phases. This model is important by itself, since it may be interpreted as a prototype of more involved entanglement phases when the clusters are endowed with lattice structure, which we will elaborate on in the following sections. In order not to digress, the phase diagram of two clusters is calculated and demonstrated in Appendix \ref{appendix:example of two cluster phase diagram}. 

\subsection{\label{ssec:scramble_hiD}Scrambling in higher dimensions}

In addition to the two-cluster interacting model, it is more interesting to consider clusters endowed with some lattice structure with nearest-neighbor interaction. Such construction creates a venue for various spin-chain many-body effects to find a new realization in the entanglement dynamics. 

As a basic example, we take the interaction to be nearest-neighbor joint random unitary evolution $\lind_U^{(1,1)}$, the form of which in large-$d$ limit is presented in Table \ref{table:summary}. We take $\mathcal{J} = d^{-4} N^{-1} $ to simplify the overall coefficient. In this example we shall see with explicit calculation that the steady-state entanglement entropy is a volume-law Page curve. The leading order takes the form of ferromagnetic Ising coupling: $\lind_U^{(1,1)} \propto -z_i z_j$. But for reasons that will come clear soon, we need to include at least the $O(d^{-1})$ term:
\begin{equation}
    \lind/N = -\sum_{\langle ij\rangle} \left( \frac{1}{2} z_i z_j + \frac{1}{2}d^{-1}(x_i+x_j) \right) .
\end{equation}
This is the transverse-field Ising model if $N=2S=1$, and it is known that when $d$ is large the ground states is doubly degenerate \cite{sachdev_2011}. We will show that it is also doubly degenerate in the large-$N$ limit through a simpler classical analysis. Minimizing over the classical variables, we get two translationally invariant ground states: $x_i = \sin\alpha$, $z_i = \pm\cos\alpha$, where $\alpha\approx 1/d$. By the same spirit as in Sec.~\ref{ssec:page}, the quantum ground states correspond to spin coherent states pointing in a  proper direction. Moreover, we assume the initial state is pure so that the final steady state is an equal-weight superposition of the two spin coherent states. So the expression for the steady state purity is
\begin{equation}
    \ket{\Psi_f} \equiv \mathcal{N}^{-1} e^{\chi\sum_iS_i^x}\cdot [e^{-i\alpha\sum_iS_i^y}|\cdots\uparrow\uparrow\uparrow\cdots\rangle+e^{i\alpha\sum_iS_i^y}|\cdots\downarrow\downarrow\downarrow\cdots\rangle]    ,
\end{equation}
\begin{equation}
    \begin{aligned}
        P_A(t=\infty)
        &= \left[ \bigotimes_{i\notin A}\bra{\uparrow_i} \cdot \bigotimes_{i\in A}\bra{\downarrow_i} \right] \ket{\Psi_f}  \\
        &= \bra{\downarrow}e^{\chi S^x}e^{-i\alpha S^y} \ket{\uparrow}^{|A|}\bra{\uparrow}e^{\chi S^x}e^{-i\alpha S^y} \ket{\uparrow}^{V-|A|}   \\
        &\phantom{=} + \bra{\downarrow}e^{\chi S^x}e^{i\alpha S^y} \ket{\downarrow}^{|A|}\bra{\uparrow}e^{\chi S^x}e^{i\alpha S^y} \ket{\downarrow}^{V - |A|}
    \end{aligned} 
\end{equation}
where $V$ is the total number of lattice points. To the leading order of $d^{-1}$, the matrix elements in the above derivation are approximated by
\begin{equation}
    \bra{\uparrow} e^{\chi S^x} e^{-i\alpha S^y} \ket{\uparrow}
    \approx \bra{\downarrow} e^{\chi S^x} e^{i\alpha S^y} \ket{\downarrow}
    \approx 1   ,
\end{equation}
\begin{equation}
    \begin{aligned}
        \bra{\uparrow}e^{\chi S^x}e^{i\alpha S^y} \ket{\downarrow}
        &\approx \prod\limits_{m'=-S}^{S-1}\sqrt{S(S+1)-m'(m'+1)}\cdot\sum\limits_{m=0}^{2S}\frac{(\alpha/2)^m}{m!}\frac{(\chi/2)^{2S-m}}{(2S-m)!} \\
        &= \left( \frac{\chi+\alpha}{2} \right)^{2S}  
        = d^{-N}    ,
    \end{aligned}
\end{equation}
and similarly, $\bra{\downarrow}e^{\theta S^x}e^{-i\alpha S^y} \ket{\uparrow} = \bra{\uparrow}e^{\chi S^x}e^{i\alpha S^y} \ket{\downarrow} = d^{-N}$ by symmetry. We finally arrive at the entanglement entropy:
\begin{equation}
    S_A(t=\infty) = \log \frac{d^{N(V- |A|)} + d^{N|A|}}{d^{NV}+1} .
\end{equation}
If we had only preserved terms up to $O((d^{-1})^0)$ order, the entropy would proportional to $\log d$, giving infinity which does not reflect the Page curve behavior. To this end, we have derived that a lattice model in the scrambled phase also has volume-law Page-curve entropy, and it resembles the entropy of a global random state. This shows that a globally scrambled state can be approached through a series of local unitary gates.

\subsection{\label{ssec:su2}Emergent SU(2) symmetry}

\paragraph{The measurement-only model}

In this subsection, we will study a more interesting situation, which is the $(1,1)$-body measurement-only model. This model exhibits an intriguing $\mathrm{SU}(2)$ symmetry (see Table \ref{table:summary}). We start with the Lindbladian expression of this model:
\begin{equation}    \label{eq:lind_su2}
    \lind/N = \sum_{\langle ij\rangle}-x_ix_j-z_iz_j+y_iy_j-2d^{-1}(x_i+x_j)-d^{-2}(x_ix_j+y_iy_j-z_iz_j)   ,
\end{equation}
where we used Eq.~(\ref{aeq:lindM_2cluster}) for $\lind_M^{(1,1)}$, and set $\lambda = 2(d^2+1)^2/N$ for notational simplicity. The inter-cluster measurements are present on nearest-neighbor pairs of clusters denoted by $\langle ij \rangle$. As a reminder, $N$ is the size of each cluster, and $L$ is the spatial length (number of clusters in 1D case). The expression Eq.~\eqref{eq:lind_su2} is exact, without going to large-$N$ or large-$d$ limits (see Appendix \ref{Appendix:two cluster lindblad} and remarks under Eq.~(\ref{aeq:leading_order})). Expanding in powers of $d^{-1}$, one immediately sees that the $O((d^{-1})^0)$ terms resemble the Heisenberg interaction, except they have staggering signs. These signs can be cured when the lattice admits bipartite structure. We are essentially going to use the identities
\begin{equation}
    e^{i\pi S^y} x e^{-i\pi S^y} = -x ,\quad
    e^{i\pi S^y} y e^{-i\pi S^y} = y   ,\quad
    e^{i\pi S^y} z e^{-i\pi S^y} = -z.
\end{equation}
For example, in a 1D chain, we perform the following rotation:
\begin{equation}
    R = e^{i\pi \sum_{i\textrm{ odd}} S^y}  ,
\end{equation}
\begin{equation}
    R \lind R^{-1} = N \sum_{\langle ij \rangle} \left( x_i x_j + y_i y_j + z_i z_j \right) + O(d^{-1})   . 
\end{equation}
We see from above that the sign difference can be resolved whenever the lattice is bipartite. In the following we will focus on this case, and leave the discussion of non-bipartite lattice (such as a 2D triangular lattice) to future works. In a 1D chain, we conclude that the problem is equivalent to the antiferromagnetic (AFM) Heisenberg model at the leading order of $d^{-1}$. One can see here why the $\mathrm{U}(1)$ symmetry in $\lind_M^{(2)}$ is enhanced to an $\mathrm{SU}(2)$ symmetry in $\lind_M^{(1,1)}$. The frustration-free nature of the lattice is essential for this symmetry enhancement. Moreover, from the perspective of the original random circuit problem, any underlying symmetry is difficult to discern due to the randomness of the evolution. It is only through the mapping to a spin model that the intricate symmetries of the original Brownian problem become manifest.

In addition, we see from Eq.~\eqref{eq:lind_su2} that the $\mathrm{SU}(2)$ symmetry is exact only in the large-$d$ limit. The terms proportional to $d^{-1}$ serve as a Zeeman field in $\hat{x}$ direction (or staggered Zeeman field after the rotation). 

\paragraph{Path integral formalism for the steady state}

\begin{figure}[t]
    \centering
    \includegraphics[width=0.54\textwidth]{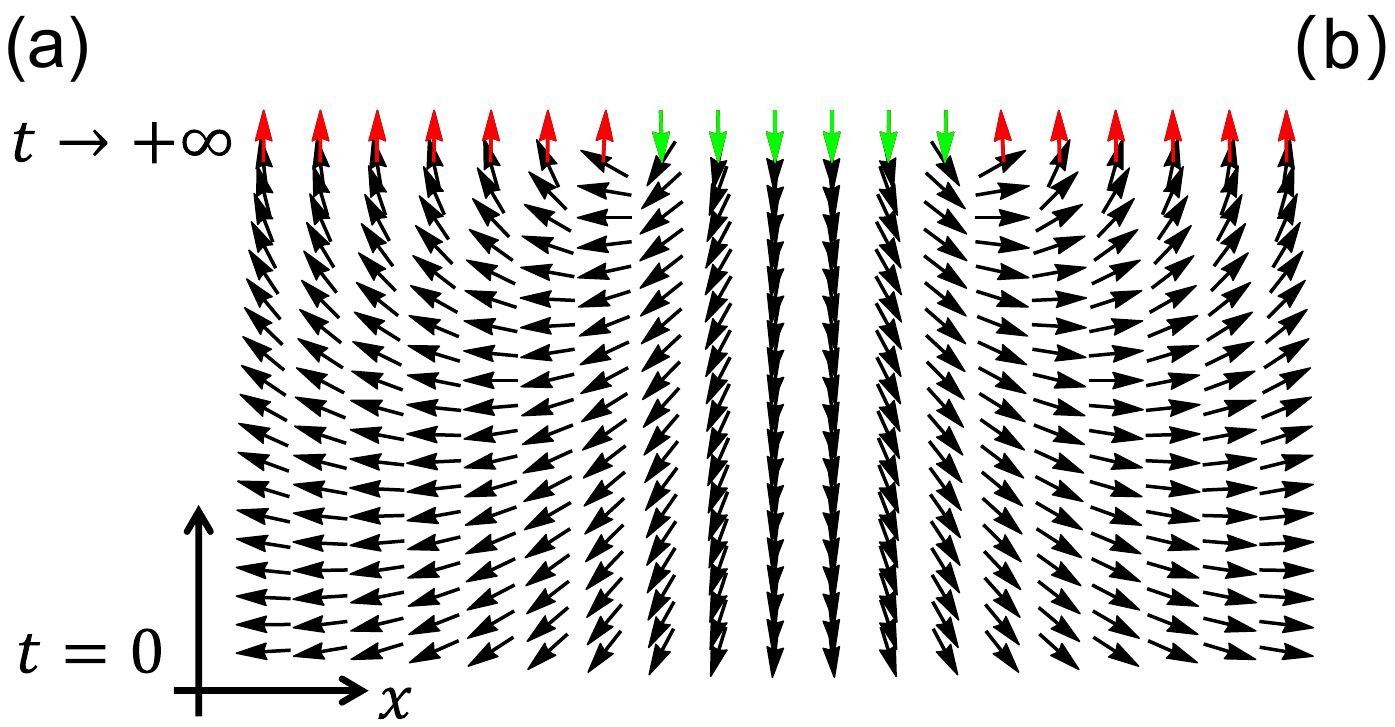}
    \includegraphics[width=0.45\textwidth]{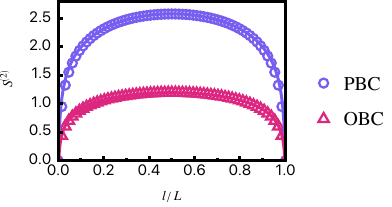}
    \caption{\label{fig:vertex illustration and DMRG} (a) Illustration for the path integral formalism. This figure represents $P_A(t=\infty)$, where $A$ is the green interval. The domain wall configuration on the boundary at $\tau=+\infty$ generates a pair of vortices, from which we infer the entropy is logarithmic with respect to interval length. (b) The steady-state second Rényi entropy $S^{(2)}$ as a function of interval length $l$. The model in this figure is Eq.~\eqref{eq:lind_su2} with $d\to \infty$, $N=2S=1$, $L=128$. The data is from a density matrix renormalization group calculation preserving bond dimension up to $200$. 
    The entropy curves are in good agreement with a $\text{CFT}_2$ entropy curves:  $S^{(2)}_{\text{PBC}} = \frac{1}{2} \log ( \frac{L}{\pi} \sin \frac{\pi l}{L} )$ and $S^{(2)}_{\text{OBC}} = \frac{1}{4} \log ( \frac{L}{\pi} \sin \frac{\pi l}{L} )$, ploted as solid lines.}

    %The entropy curves are in good agreement with a $\text{CFT}_2$ entropy with central charge $c=2$ \cite{CFT}. We plot the CFT results $S^{(n)}_{\text{PBC}} = \frac{c}{6}(1+\frac{1}{n}) \log ( \frac{L}{\pi} \sin \frac{\pi l}{L} )$ and $S^{(n)}_{\text{OBC}} = \frac{c}{12}(1+\frac{1}{n}) \log ( \frac{L}{\pi} \sin \frac{\pi l}{L} )$ as solid lines, with the Rényi index $n=2$. }
\end{figure}

We proceed with studying the steady state(s) that the system converges to in the infinite time limit. The purity $P_{A}(t=\infty)$ ($A$ is a subsystem when calculating the entanglement entropy) is determined by the ground state of $\lind$, overlapping with a specific state corresponding to $A$. This problem admits a path integral formalism in terms of spin coherent states. A picture of this idea is shown in Fig.~\ref{fig:vertex illustration and DMRG}(a). The derivation of the action is standard \cite{book2}, resulting in a non-linear sigma model (NLSM) with Berry phase:
\begin{equation}
    \int\mathcal{D}\hat{\Omega}\, e^{i\Upsilon[\hat{\Omega}]} \exp \left[ -\frac{S}{4} \int dxdt\, (\partial_x\hat{\Omega})^2 + (\partial_t\hat{\Omega})^2 \right]    .
\end{equation}
The explicit expression of the Berry phase is
\begin{equation}
    \Upsilon[ \hat{\Omega}(x,t) ]
    = 2\pi S \cdot \Theta[ \hat{\Omega}(x,t) ]
    = 2\pi S \cdot \frac{1}{4\pi} \int dxdt\, (\hat{\Omega} \times \partial_t \hat{\Omega}) \cdot \partial_x \hat{\Omega} ,
\end{equation}
where $\Theta$ is the Pontryagin index, counting the winding number of $\hat{\Omega}$ as a mapping from $t>0$ to the unit sphere. Then we need to specify the boundary conditions of for this path integral. Since the ground state is concerned, the time interval of the path integral is infinite, and the domain of integration is $t>0$. The initial condition is set at $t=0$, and the final condition is at $t=+\infty$, determined by the subset $A$. Remember from Sec.~\ref{sec:setup} that the purity is $P_A(\infty) = \pbra{X_A} e^{-\lind \tau} \pket{\sigma^{\otimes 2}}$. In spin language, $\pbra{X_A}$ translates into spin $S^z_i=S=N/2$ when $i\notin A$, and $S^z_i=-S=-N/2$ when $i\in A$\footnote{Here we are only dealing with the case that each cluster is treated as a whole, i.e., either completely in or not in the set $A$. Generally, one can also design $A$ to cover a part of a cluster, say $n$ qudits out of $N$, and spin boundary condition should be $\bra{S^z=S-n} \mathcal{N}^{-1} e^{\chi S^x}$. }. Also, an operator $e^{\chi S^x}$ need to be inserted to compensate the pseudo-Hermitian transformation (see Eq.~\eqref{eq:pur_vec_evo}), but it is asymptotically trivial in the large-$d$ limit (recall $\tanh\chi=d^{-1}$). Therefore, the final state for the path integral is 
\begin{equation}
    \bigotimes_{i\notin A}\bra{\uparrow_i} \cdot \bigotimes_{i\in A}\bra{\downarrow_i} .
    % e^{\chi\sum_i S_i^x} .
\end{equation}
This is also illustrated in Fig.~\ref{fig:vertex illustration and DMRG}(a), in which the green interval stands for $A$. The entanglement entropy of a subsystem $A$ is thus the free energy difference between the illustrated vortex configuration and the homogeneous spin up configuration. Especially, in the large-$N$ limit the free energy is dominated solely by the saddle-point configuration fixed by the boundary condition (see Fig.~\ref{fig:vertex illustration and DMRG}(a)). 

Having the path integral interpretation in mind, we can discuss some representative cases. For simplicity, we restrict our discussion to periodic boundary case and assume $L$ is even.  
\begin{enumerate}
    \item \emph{Infinite $N$.} When $N$ approaches infinity, the NLSM action is multiplied by a large factor and therefore all quantum fluctuations are suppressed. We can focus on the classical saddle point configurations, which suggests that the domain walls on the boundary generate vortices in the bulk (see Fig.~\ref{fig:vertex illustration and DMRG} (a)), so the entropy is just the free energy of vortex interaction,  which is logrithmic in distance. So, we expect that the entropy scales as $S \propto \log|x - y|$, where $x$ and $y$ are end points of subregion $A$. 
    \item \emph{Even $N$.} Under periodic boundary condition, the so-called Haldane gap is present between the ground state and the first excited state, which scales as $\Delta\propto e^{-\pi S}$ \cite{affleck1989}. We see that the Haldane gap vanishes when $N\to\infty$, which agrees with the previous infinite-$N$ analysis. The entanglement entropy is volume-law due to the finite gap. 
    
    % Particularly, when $N=2$ the ground state exhibits symmetry-protected topological (SPT) order, belonging to the same class as the Affleck-Kennedy-Lieb-Tasaki (AKLT) state. 
    % On the other hand, under open boundary condition and $N=2$, we would expect SPT edge states to appear, which are responsible for interesting four-fold degenerate ground states. 
    \item \emph{Odd $N$.} The low-energy physics is drastically different in a half-integer spin chain. As a consequence of the Lieb-Schutz-Mattis theorem, half-integer spin chain has a non-degenerate ground state and gapless low-energy excitations \cite{affleck1989}. Consequently, the two-point correlation functions exhibit power-law decay, indicating that the entropy scales as $S \propto \log|x - y|$, where $x$ and $y$ are end points of subregion $A$. This behavior is supported by the numerical study of the $N = 1$ case shown in Fig.~\ref{fig:vertex illustration and DMRG}(b). Moreover, gaplessness implies that the relaxation towards the steady state follows an algebraic, rather than exponential, decay.
\end{enumerate}

It is particularly surprising to see a conformal field theory (CFT) entropy curve in random circuits in Fig.~\ref{fig:vertex illustration and DMRG} (b), in which we took $N=2S=1$. Beside what has been plotted, the entropy in fact exhibits an even-odd oscillation. In terms of the purity, the purity is finite at even $l$, while is zero at odd $l$, which results in a divergent entropy. This can also be understood through an entanglement pair argument similar to that at the end of Sec.~\ref{ssec:u1}. Due to the divergence at odd $l$'s, we only plot the even $l$ data in Fig.~\ref{fig:vertex illustration and DMRG} (b), which form a CFT entropy curve by themselves: for periodic boundary condition, we have $S_{\text{PBC}}(l)=\frac{1}{2}\log(\frac{L}{\pi}\sin\frac{\pi l}{L})$. In large volume limit $(L\rightarrow\infty)$, we have $S_{\text{PBC}}(l)=\frac{1}{2}\log|l|$. The prefactor $\frac{1}{2}$ is related to the scaling dimension of a certain boundary-condition changing operator in Luttinger liquid boundary CFT. More details are presented in Appendix~\ref{app: Luttinger}. For OBC we have $S_{\text{OBC}}(l)=\frac{1}{4}\log|l|$, with the coefficient $\frac{1}{4}$ being half of PBC. This can be interpreted intuitively since the OBC configuration only contains half of the vortex configuration drawn in Fig.~\ref{fig:vertex illustration and DMRG}(a), so the excitation energy is half.

\paragraph{Physical steady state}

It may be puzzling why the steady-state purity is always physical, i.e., all its components are real and positive. This is guaranteed by Marshall's theorem for spin chains \cite{book2}. Recalling that $P_n(t=\infty) = \bra{n} \mathcal{N}^{-1} e^{\chi\sum_i S_i^x} \ket{\Psi_0} $, the positivity of the purity directly translates into the positivity of $\bracket{n}{\Psi_0}$ itself, where $\ket{\Psi_0}$ is the ground state. Marshall's theorem states that for the AFM Heisenberg model on a bipartite lattice, the ground state wave function in $S^z$ eigenbasis has the following sign structure: 
\begin{equation}
    \operatorname{sgn} \bracket{m_1 m_2 \dots m_L}{\Psi_0} = (-1)^{\sum_{i\text{ odd}} (S+m_i)} ,
\end{equation}
where the bra state is an $S^z$ eigenbasis of the whole spin chain, with $m_i = -S, -S+1, \dots , S$, $i = 1, 2, \dots , L$. In our specific model, the rotated Lindbladian is exactly the AFM Heisenberg model and thus has the above ground state wave function. But the rotation $\exp \left( i\pi\sum_{i\text{ odd}} S_i^y \right)$ exactly cancels Marshall's signs, so we conclude that the ground state of Eq.~\eqref{eq:lind_su2} has uniform sign, indicating that the steady state purity is always physical. 

In general cases of finite $d$, we can prove that the ground state is real and positive from another angle. We need the following simple lemma. Let $A$ be a real symmetric matrix and all its off-diagonal elements are non-positive. If $A$ has a unique ground state, its wave function must have uniform sign. Since in Eq.~\eqref{eq:lind_su2} we have off-diagonal terms $-x_ix_j+y_iy_j\propto-(S_i^+S_j^++S_i^-S_j^-)$, $-x_ix_j-y_iy_j\propto-(S_i^+S_j^-+S_i^-S_j^+)$, and $-(x_i+x_j)\propto-(S_i^++S_i^-+S_j^++S_j^-)$, which are all element-wise negative, the lemma applies to our Lindbladian and therefore the purity ground state is always physical. 

The lemma is proved as follows. The ground state problem amounts to the variational minimization of $\bra{\Psi} A \ket{\Psi}$, where $A$ is a $D\times D$ matrix. We expand it as
\begin{equation}
    \bra{\Psi} A \ket{\Psi} = \sum_{a=1}^D (A)_{aa} \rho_a^2 + 2\sum_{a<b} (A)_{ab} \rho_a \rho_b \cos(\theta_a-\theta_b)   ,
\end{equation}
where we introduced a decomposition for the wave function into its modulus and angle: $\Psi_a = \rho_a e^{i\theta_a}$, $a=1,2,\dots , D$. Because $(A)_{ab}\leq 0$, the expectation is minimized only when all $\theta_a$'s are identical, so they can be set to zero identically.

Finally, we remark on possible alternation to this $\mathrm{SU}(2)$ model. The cluster chain can also exhibit entanglement phase transition by tuning parameters. For example, we add single-body measurement on each site, and the purified or the scrambled phase is expected to appear according to the relative strength of the two kinds of measurements, $\lind_M^{(1)}$ and $\lind_M^{(1,1)}$.  

\section{\label{sec:conclusion}Conclusion and discussions}
% \paragraph{Open questions and interesting directions}

In the current work, an analytic formulation for entanglement purity evolution is established. We constructed a mapping from random-averaged quantum entanglement evolution to imaginary-time evolution of spin many body model. In the limit of large-$N$ (cluster size), the spins admit a semiclassical treatment, which makes numerous concrete random circuit models analytically tractable. Our formulation is exploited to understand the standard measurement-induced phase transitions, as well as entanglement phase transitions in measurement-only circuits. In measurement-only models, our formulation is able to predict quantitative bound for the measurement locality $q'$ that is necessary for a globally scrambled phase. Furthermore, in the large-$d$ (qudit dimension) limit, we discoverd emergent $\mathrm{U}(1)$ symmetry in all-to-all random measurement circuits, and emergent $\mathrm{SU}(2)$ symmetry in circuits with spatial structure. The emergent symmetries are responsible for a critical point between the usual scrambled phase and the purified phase, which we call the critical purification. The critical point is characterized by two properties. First, the final steady state is unique, but the entanglement entropy is super-area-law. Second, the purification time (typical time needed to approach the steady state) grows with the size of the system. 

%Our new understanding on measurement circuits may provide a brickwork towards measurement-based quantum computation \cite{NIELSEN200396,Briegel2009}. \XLQc{Could you elaborate the relation with MBQC}\HYW{Instead of MBQC (although of course SPT states are MBQC resource states), I am thinking of the following outlook. }

Moreover, our work advances the understanding on measurement-based state preparation protocols. There has been an increasing interest in preparing highly entangled states with measurements. The most prominent examples are the toric code ground state and 1D SPT states---both can be prepared within finite depth of projective measurements, but require deep circuits if prepared only by unitary gates (see e.g., \cite{PRXQuantum.3.040337, PhysRevLett.127.220503, Iqbal2024}). Our work analyzes the entanglement-generating ability of \emph{random} projective measurement, thus shedding light on the \emph{typical} cost of preparing useful entangled states via measurement. 

Finally, we propose some future directions related to the current work. 

\begin{itemize}
\setlength{\parskip}{0pt}  
\item
\textbf{Field theory near criticality. }Near the critical point of entanglement phase transition, an effective field theory can be written down as introduced in Appendix \ref{app:gap}. This opens up the way to understand the entanglement phase transitions in Ginzburg-Landau paradigm. In higher spatial dimensions, similar derivation also leads to a field theory description near criticality, which is worth exploring in future works. The quadratic terms in $\Pi$ and $\Phi$ will turn into non-interacting parts of the Hamiltonian, while interaction can be included by expanding $\lind$ up to quartic terms. 
\item
\textbf{Topological phases and structured circuits. }Beside conventional Ginzburg-Landau phase transitions, topological phase transitions are also feasible in our formulation of random circuit models. In Sec.~\ref{ssec:su2}, we have encountered the simplest symmetry-protected topologically non-trivial state, i.e., the AKLT state in the $\mathrm{SU}(2)$ measurement-only model with $N=2$. Under open boundary condition, the ground state degeneracy is four. More generally, it is possible to construct topological ordered states, such as quantum spin liquid on a 2D lattice. These interesting combinations between entanglement phase and topological order are left for future investigation. 
\item
\textbf{Fractal entanglement in long-range coupled models. }A fractal entanglement phase was reported in Ref. \cite{Open3}, featuring an entanglement entropy scaling as $S\propto L^{\alpha}$ with non-integer exponent $\alpha$ . The fractal entanglement phase arises in a non-unitary free fermion chain model with power-law long-range interaction. The formalism depends on the quadratic Hamiltonian of free fermions, making it hard to generalize to interacting systems, e.g., random circuits or SYK chains. However, our large-spin interpretation of the entanglement dynamics admits a quadratic Hamiltonian when expanded near critical point (see Appendix \ref{app:gap}). Therefore, our work suggests that a large qudit cluster may be another platform for realizing the fractal entanglement phase. 
\end{itemize}

% \emph{Note added: }Upon finishing this manuscript, we became aware of two papers that coincide with our idea of measurement-induced emergent symmetries \cite{jian2023measurement,fava2023nlsm}. However, the authors discovered emergent symmetries and a non-linear sigma model in free-fermion models, while we based on a distinct underlying physical system of qudits. 

\section*{Acknowledgements} 

HT and HYW would like to thank Tian-Gang Zhou and Chengshu Li for discussion on the interpretation of the Weingarten functions. 
% HT would like to thank Zhiyuan Yao for pointing out the nonlinear sigma model as dual of quantum antiferromagnets, and Changyan Wang for bringing attention to Marshall's theorem. 
HT would also like to thank Pengfei Zhang, Yi-Zhuang You, Zhiyuan Yao and Changyan Wang for helpful discussions. HT especially thanks Professor Hui Zhai for his  support and encouragement throughout this work. HYW thanks Zijian Wang for helpful discussion. The DMRG calculation was carried out using the TeNPy Library (version 0.10.0) \cite{tenpy}. This work proceeded partially during XLQ's visit to the Institute for Advanced Study, Tsinghua University (IASTU). XLQ is grateful for the hospitality of IASTU. ZW is supported by the NSFC under Grant No. 12125405. XLQ is supported by the Simons Foundation. HT is supported by Shoucheng Zhang fellowship.

\appendix
% appendix.tex

\section{Derivation of Lindbladian on a single cluster}
\label{Appendix:one cluster lindblad}

In this Appendix, we elaborate on the matrix representation of the Lindbladian corresponding to $q$-body unitary evolution and $q'$-body projective measurement. These terms are denoted as $\tilde{\lind}_U$ and $\tilde{\lind}_M$, respectively. The basic construction of these terms are introduced in Sec.~\ref{ssec:map_classical}. 

\subsection{Derivation of $\tilde{\lind}_U$}
We first derive $\tilde{\lind}_U$ for the random unitary evolution part. We restate its definition here. $\tilde{\lind}_U$ originates from the operator $\mathscr{L}_U$:
\begin{equation}    \label{aeq:def_LU}
    \mathbbm{1} - \mathscr{L}_U \Delta t = \overline{ \mathcal{T}\exp \left[ -i \int_t^{t+\Delta t} dt' \left( H^1(t') - H^{\bar{1}*}(t') + H^2(t') - H^{\bar{2}*}(t') \right) \right] }    ,
    % \overline{ e^{-iH(t)\Delta t} \otimes e^{iH^*(t)\Delta t} \otimes e^{-iH(t)\Delta t} \otimes e^{iH^*(t)\Delta t} }   .
\end{equation}
where $1$, $\bar{1}$, $2$, $\bar{2}$ are labels for the replicas. $\tilde{\lind}_U$ is the matrix representation defined as
\begin{equation}
    \pbra{X_A} \mathscr{L}_U = \sum_{B} (\tilde{\lind}_U)_{AB} \pbra{X_B}  ,
\end{equation}
where $A$ and $B$ are subsets of the whole system. 

Through Taylor expansion in Eq.~(\ref{aeq:def_LU}), the first nontrivial term $-i\overline{\int^{\Delta t}dt'\, H}$ averages to zero. But the quadratic term in $H(t)$ does not vanish, and is actually first-order in $\Delta t$:
\begin{equation}    \label{aeq:lind_average}
    \begin{aligned}
        &\phantom{=} \mathscr{L}_U\Delta t \\
        &= \int_{t}^{t+\Delta t} dt' \int_{t}^{t'} dt'' \overline{ \left( H^1(t') - H^{\bar{1}*}(t') + H^2(t') - H^{\bar{2}*}(t') \right) \cdot \left( t'\to t'' \right) }    \\
        &= \frac{1}{2} \sum_{\mathbf{i}, \mathbf{a}} \int_t^{t+\Delta t} dt'\, \mathcal{J} \left( T_{i_1 a_1}^{1} \cdots T_{i_q a_q}^{1} - T_{i_1 a_1}^{\bar{1}*} \cdots T_{i_q a_q}^{\bar{1}*} + T_{i_1 a_1}^{2} \cdots T_{i_q a_q}^{2} - T_{i_1 a_1}^{\bar{2}*} \cdots T_{i_q a_q}^{\bar{2}*} \right)^2 \\
        &= \frac{1}{2} \mathcal{J} \Delta t \sum_{\mathbf{i}, \mathbf{a}} \left( T_{i_1 a_1}^{1} \cdots T_{i_q a_q}^{1} - T_{i_1 a_1}^{\bar{1}*} \cdots T_{i_q a_q}^{\bar{1}*} + T_{i_1 a_1}^{2} \cdots T_{i_q a_q}^{2} - T_{i_1 a_1}^{\bar{2}*} \cdots T_{i_q a_q}^{\bar{2}*} \right)^2    ,
    \end{aligned}
\end{equation}
where $\sum_{\mathbf{i}, \mathbf{a}}$ is a shorthand for $\sum_{1\leq i_1<\dots<i_q \leq N}\sum_{a_1, \dots, a_q}$. One should recall here Eq.~(\ref{eq:Brownian_ham}), the definition of the Hamiltonian, which contains a Brownian coefficient. Then, summing over the Hermitian operators $T$, we arrive at these identities:
\begin{align}
    \sum_a T_{ia}^1 T_{ia}^1 &= d^2 , \\
    \sum_a T_{ia}^1 T_{ia}^2 &= dX^{12}_i=d\sum\limits_{a,b=1}^{d}(|a\rangle\langle b|)_i\otimes \mathbbm{1}_i\otimes(|b\rangle\langle a|)_i\otimes \mathbbm{1}_i   ,   \\
    \sum_a T_{ia}^1 T_{ia}^{\bar{1}*}&=dP^{1\bar{1}}_i=d\sum\limits_{a,b=1}^{d}(|a\rangle\langle b|)_i\otimes(|a\rangle\langle b|)_i\otimes \mathbbm{1}_i \otimes \mathbbm{1}_i  .
\end{align}
Here, $X^{12}$ refers to the SWAP operator between replicas $1$ and $2$; $P^{1\bar{1}}$ refers to a projector onto the maximally entangled state between $1$ and $\bar{1}$. 

Next, we aim to find a matrix representation for $\mathscr{L}_U$. $\pbra{X_A}$ can be understood as a SWAP state on subsystem $A$. Acting on the left vectors $\pbra{I_i^{\pm}}$, we have the following matrix representations for relevant operators:
\begin{align}
    \begin{bmatrix}
        (I^+_i| \\ (I^+_i|
    \end{bmatrix}
    X^{12}_i \text{ (or } X^{\bar{1}\bar{2}}_i \text{)} &=
    \begin{bmatrix}
        0\  & \ 1 \\
        1\  & \ 0 
    \end{bmatrix}
    \begin{bmatrix}
        (I^+_i| \\ (I^+_i|
    \end{bmatrix}
    ,\\
    \begin{bmatrix}
        (I^+_i| \\ (I^+_i|
    \end{bmatrix}
    P^{1\bar{1}}_i \text{ (or } P^{2\bar{2}}_i \text{)} &=
    \begin{bmatrix}
        d\  & \ 0 \\
        1\  & \ 0 
    \end{bmatrix}
    \begin{bmatrix}
        (I^+_i| \\ (I^+_i|
    \end{bmatrix}
    ,\\
    \begin{bmatrix}
        (I^+_i| \\ (I^+_i|
    \end{bmatrix}
    P^{1\bar{2}}_i \text{ (or } P^{2\bar{1}}_i \text{)} &=
    \begin{bmatrix}
        0\  & \ 1 \\
        0\  & \ d 
    \end{bmatrix}
    \begin{bmatrix}
        (I^+_i| \\ (I^+_i|
    \end{bmatrix}
    .
\end{align}
It is a crucial simplification to our problem that the SWAP operators and the projectors are closed in the subspace spanned by $\pbra{I_i^+}$ and $\pbra{I_i^-}$. We have interpreted these two bases as spin up and spin down, respectively, in the main text. Also, we recall that $\tilde{\lind}_U$ is by definition the matrix representation of $\mathscr{L}_U$ in the space spanned by $\pbra{I_i^{\pm}}$. Denoting the matrices on the right-hand side by $A$, $B$, and $C$, respectively, we arrive at
\begin{equation} \label{aeq:lindU_exact}
    \tilde{\lind}_U
    = 2\mathcal{J} \sum_{i_1<\dots<i_q}(d^{2}\dots d^{2} + dA_{i_1}\dots dA_{i_q} - dB_{i_1}\dots dB_{i_q} - dC_{i_1}\dots dC_{i_q})    ,
\end{equation}
\begin{equation}
    A_i=2S^x_i,\ B_i=d/2+S^x_i-iS^y_i+dS^z_i,\ C_i=d/2+S^x_i+iS^y_i-dS^z_i  ,
\end{equation}
where $S_i^{x,y,z} = \sigma_i^{x,y,z} /2$ are spin-half matrices acting on site $i$. Due to the equal role played by each spin, we are able to write $\tilde{\lind}_U$ solely in terms of the total spin operators. In the large-$N$ limit, we may treat $S^{x,y,z}_{\mathrm{tot}} = \sum_i S^{x,y,z}_{\mathrm{tot}}$ as $O(N)$, and preserve only the leading order in $N$ as
% Notice that since $L_{0U}$ has permutation symmetry, it can be expressed solely in terms of total spin operator. This process can be explicitly carried out if we work in large $N$ limit:
\begin{equation}    \label{aeq:largeN_approx}
    \sum_{i_1<\dots<i_q}A_{i_1}\cdots A_{i_q}
    \approx \frac{1}{q!} \left(\sum_iA_i\right)^q
    = \frac{1}{q!} (A_{\mathrm{tot}})^q ,
\end{equation}
and similar equations hold for $B$ and $C$. Therefore, 
\begin{equation}
    \tilde{\lind}_U = \frac{2\mathcal{J}}{q!} \left[ (d^2N)^q + (dA_{\mathrm{tot}})^q - (dB_{\mathrm{tot}})^q - (dC_{\mathrm{tot}})^q \right] + O(\mathcal{J} N^{q-1})  .
\end{equation}
Putting the expressions of $A$, $B$ and $C$ into the above formula, we finally have the matrix representation of $\tilde{\lind}_U$ in large-$N$ limit:
\begin{equation}    \label{aeq:lindU}
    \begin{aligned}
        \tilde{\lind}_U = \frac{2\mathcal{J}d^q}{q!} \Big[ (dN)^q &+ (2S^x_{\mathrm{tot}})^q 
        - \left(\frac{d}{2}N+S_{\mathrm{tot}}^x-iS_{\mathrm{tot}}^y+dS_{\mathrm{tot}}^z\right)^q   \\
        &- \left(\frac{d}{2}N+S_{\mathrm{tot}}^x+iS_{\mathrm{tot}}^y-dS_{\mathrm{tot}}^z\right)^q \Big] + O(\mathcal{J} N^{q-1})   .
    \end{aligned}
\end{equation}

\subsection{Derivation of $\tilde{\lind}_M$}

\begin{figure}[t]
    \centering
    \includegraphics[width=1\textwidth]{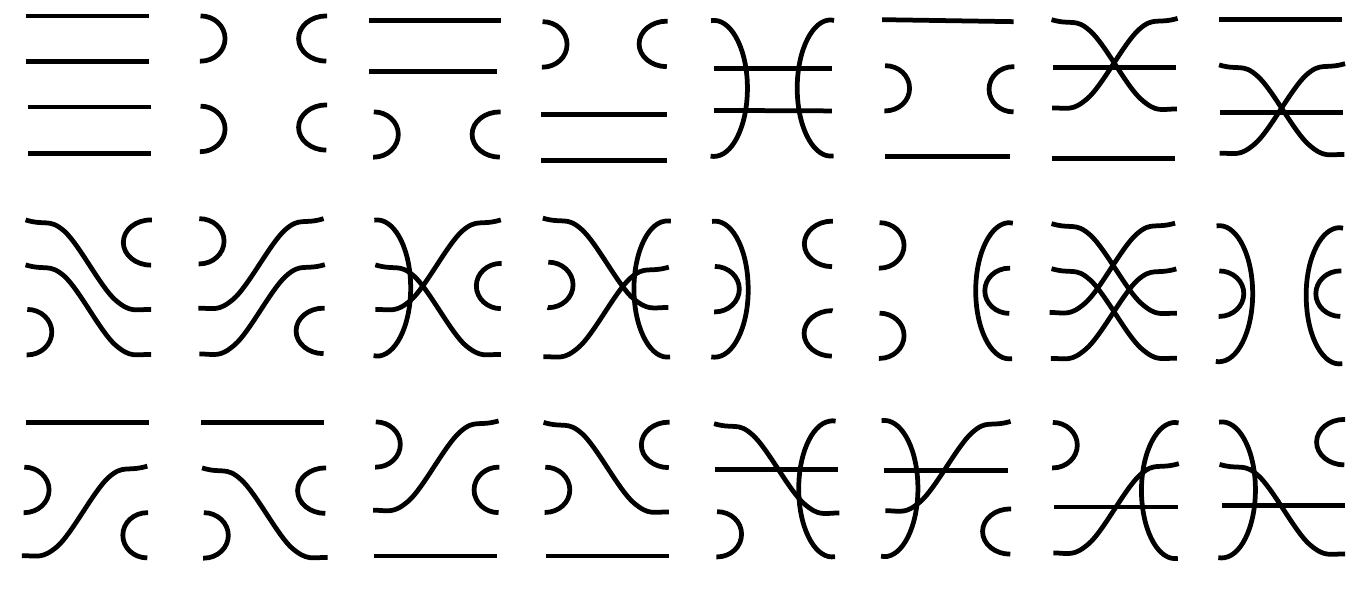}
    \caption{\label{fig:Weingarten}Diagrams of summands due to the Weingarten functions. These are all of the $4! =24$ permutations in the four-replica Hilbert space. For example, the diagram in the second row, third column means $\sum\limits_{a,b,c,e=1}^{d}|acea\rangle\langle ebbc|$ explicitly. Their sum times $\mathrm{Wg}_0$ gives $M$ in Eq.~(\ref{aeq:wgsum}). } 
\end{figure}

Here we derive the Lindbladian in the replicated space due to random projective measurements. In deriving $\tilde{\lind}_M$, the basic expression we need to evaluate is
\begin{equation}
    \begin{aligned}
    M\equiv\overline{|\varphi\varphi^*\varphi\varphi^*\rangle\langle\varphi\varphi^*\varphi\varphi^*|}&=\overline{U\otimes U^*\otimes U\otimes U^*|0000\rangle\langle0000|U^{\dagger}\otimes U^{\dagger*}\otimes U^{\dagger}\otimes U^{\dagger*}}\\
    &\longrightarrow\overline{U\otimes U^*\otimes U\otimes U^*\otimes U^*\otimes U\otimes U^*\otimes U}|0000\rangle|0000\rangle ,
    \end{aligned}
\end{equation}
where $U$ is a random matrix drawn from Haar measure. In the second line, we mapped the operator into a state. Hence here an eight-$U$ average need to be done with help of the Weingarten functions \cite{Collins2003}. In principle, the average of eight $U$'s can involve numerous terms. But acting on a simple state $\ket{00000000}$, the result is greatly simplified to an equal-weight sum over all permutation operators, with an over all coefficient $\text{Wg}_0\equiv\sum_{\sigma}\text{Wg}(\sigma)=(D^4+6D^3+11D^2+6D)^{-1}$, where $D$ is the Hilbert space dimension of $\ket{\varphi}$. These terms are shown in Fig.~\ref{fig:Weingarten}. Also, all these permutation operators are closed in the 2D subspace spanned by $\pbra{I^{\pm}}$, whose matrix elements read
\begin{equation}    \label{aeq:wgsum}
    \begin{bmatrix}
        (I^+| \\ (I^- |
    \end{bmatrix}
    M=\text{Wg}_0
    \begin{bmatrix}
        D^2+5D+6\  & \ D^2+5D+6\\
        D^2+5D+6\  & \ D^2+5D+6
    \end{bmatrix}
    \begin{bmatrix}
        (I^+| \\ (I^-|
    \end{bmatrix}
    =\frac{1}{D(D+1)}
    \begin{bmatrix}
        1\  & \ 1\\
        1\  & \ 1
    \end{bmatrix}
    \begin{bmatrix}
        (I^+| \\ (I^-|
    \end{bmatrix}   .
\end{equation}
As a remark, the effect of measurement $\overline{\ket{\varphi}\bra{\varphi}}$ here has exactly the same representation as the kind of measurement discussed in Ref.~\cite{MIPT30}. In the latter, the measurement is designed as $\overline{\ket{\phi}\bra{\varphi}}$, where $\ket{\phi}$ and $\ket{\varphi}$ are independent Haar random states. It can be interpreted as a projective measurement followed by a random unitary operator. Though their reason of using such form is to avoid complicated eight-$U$ average, we have shown that the more conventional measurement $\overline{\ket{\varphi}\bra{\varphi}}$ has identical effect. Alternatively, noticing the identities $( I^{\pm} | I^{\pm} ) = d^2$ and $( I^{\pm} | I^{\mp} ) = d$, we are able to write $M$ as a ``projector'':
\begin{equation}
    M = \pket{S} \pbra{S}   ,\quad
    \pket{S} = \frac{\pket{I^+}+\pket{I^-}}{D(D+1)} .
\end{equation}

Next, we discuss what form $\tilde{\lind}_M$ should take. It has been found in the main text that $\tilde{\lind}_M$ is related to $M$ by
% Recall that a measurement $\ket{\varphi}\bra{\varphi}$ happens at rate $\lambda$, so the effect on the duplicated density matrix is
% \begin{equation}
%     \pket{\sigma^{\otimes 2}} \to \left[ 1 + \lambda\Delta t (1-M) \right] \pket{\sigma^{\otimes 2}}    .
% \end{equation}
% We also notice that we need to sum over the choices of the $q'$ measured qudits. The dimension of state $\ket{\varphi}$ should be $D = d^{q'}$. Therefore, 
\begin{equation}
    \tilde{\lind}_M = \lambda \sum_{\mathbf{i}}\left( \mathbbm{1} - \pket{S_{\mathbf{i}}} \pbra{S_{\mathbf{i}}} \right) ,
\end{equation}
where, as in the main text, $\mathbf{i}$ is a shorthand for $1 \leq i_1 < \dots < i_{q'} \leq N$. Furthermore, we introduce the following four $2\times 2$ matrices
\begin{equation}
    \begin{bmatrix}
        \pbra{I^+_i} \\ \pbra{I^-_i}
    \end{bmatrix}
    \pket{I^{s_1}_i}\pbra{I^{s_2}_i}
    = A^{s_1s_2}
    \begin{bmatrix}
        \pbra{I^+_i} \\ \pbra{I^-_i}
    \end{bmatrix}
    \text{, }
    s_1, s_2 = \pm,
\end{equation}
\begin{equation}
    A^{++} = 
    \begin{bmatrix}
        d^2 & 0 \\ d & 0
    \end{bmatrix}
    \text{, }
    A^{--} = 
    \begin{bmatrix}
        0 & d \\ 0 & d^2
    \end{bmatrix}
    \text{, }
    A^{+-} = 
    \begin{bmatrix}
        0 & d^2 \\ 0 & d
    \end{bmatrix}
    \text{, }
    A^{-+} = 
    \begin{bmatrix}
        d & 0 \\ d^2 & 0
    \end{bmatrix}
    \text{. }
\end{equation}
Using these notations, we arrive at the precise matrix representation (here $D=d^{q'}$)
\begin{equation} \label{aeq:lindM_exact}
    \tilde{\lind}_M
    = \lambda\sum_{1\leq i_1<\dots<i_{q'} \leq N}\left(1-\frac{A^{++}_{i_1}\cdots A^{++}_{i_{q'}}+A^{--}_{i_1}\cdots A^{--}_{i_{q'}}+A^{+-}_{i_1}\cdots A^{+-}_{i_{q'}}+A^{-+}_{i_1}\cdots A^{-+}_{i_{q'}}}{(d^{q'}(d^{q'}+1))^2} \right)   .
\end{equation}

Finally, with the same spirit as in Eq.~(\ref{aeq:largeN_approx}), we may evaluate the sum effortlessly in the large-$N$ limit. The result is
\begin{equation}
    \begin{aligned}
        \tilde{\lind}_M &= \frac{\lambda}{q'! (d^{q'}+1)^2} \Big[ (d^{q'}+1)^2 N^{q'}\\
        &-\left( \frac{N}{2} + S^z + \frac{1}{d}(S^x-iS^y) \right)^{q'}
        -\left( \frac{N}{2} - S^z + \frac{1}{d}(S^x+iS^y) \right)^{q'}\\ 
        &-\left( S^x-iS^y + \frac{1}{d}(\frac{N}{2}+S^z) \right)^{q'}
        -\left( S^x+iS^y + \frac{1}{d}(\frac{N}{2}-S^z) \right)^{q'} \Big] + O(\lambda N^{q'-1})  ,
    \end{aligned}
    \label{aeq:lindM}
\end{equation}
% From the above formula, we may rescale $\lambda$ by $\frac{\lambda}{q'! (d^{q'}+1)^2} \to \lambda$ to make the expression of the Lindbladian simpler. 

We have seen that $\tilde{\lind}_U$ and $\tilde{\lind}_M$ appear to be non-Hermitian. Actually, they can be put into Hermitian form by a similarity transformation (pseudo-Hermitian property), as discussed in Sec.~\ref{ssec:pseudo-h}. 

As a remark, the Lindbladians specified to $q=2$ and $q'=1$ are reported in Ref.~\cite{MIPT30}, which conforms with our formulas, Eqs.~(\ref{aeq:lindU_exact}) and (\ref{aeq:lindM_exact}).

\section{\label{app:gap}Liouvillian gap and criticality in measurement-only models}

In this Appendix we give an estimation for the gap $\Delta$ between the ground state and the first excited state of the Lindbladian. This gap is important because it reflects the time scale needed to approach the final steady state, since the relaxation time is roughly $1/\Delta$. 

We demonstrate the calculation in all-to-all measurement-only models that have been introduced in Sec.~\ref{ssec:MIS}. Regarding the $q'$-body measurement Lindbladian $\lind_M^{(q')}$, it is known that a classical spontaneous symmetry breaking of the subsystem inversion symmetry may happen around $(x,y,z)=(1,0,0)$, or equivalently, it is around $\varphi=0$ with polar coordinates $x=\cos\varphi$, $z=\sin\varphi$. The symmetry-preserving phase corresponds to the purified phase, while the symmetry-broken phase (i.e., ground states at $\varphi = \pm\varphi_0 \neq 0$) corresponds to the scrambled phase. 

To see the $N$-dependence of the Liouvillian gap as well as the low-energy\footnote{Here, ``energy'' refers to the eigenvalues of the entanglement Lindbladian. } spectrum, we proceed with a quantum-mechanical treatment, which enriches our semiclassical treatment in the main text. Assuming the low-energy states lie in a neighborhood of $\hat{x}$-direction, we can use the Holstein-Primakoff boson representation as \cite{book2}
\begin{equation}    \label{aeq:HPrep}
    \begin{aligned}
        S^x &= S - b^\dagger b   ,   \\
        S^y + i S^z &= \sqrt{2S - b^\dagger b} b ,   \\
        S^y - i S^z &= b^\dagger \sqrt{2S - b^\dagger b} ,   \\
    \end{aligned}
\end{equation}
where $b$ is a boson annihilation operator satisfying $[b, b^\dagger] = 1$, and by our definition, $S=N/2$. The Holstein-Primakoff boson is effectively the low-energy excitation mode around the classical solution. If we want to expand $\lind_M^{(q')}$ in powers of $N$ and preserve only terms at the order of $O(N)$ and $O(1)$ (by a proper scaling of $\lambda$, $\lind_M^{(q')}$ is set to be $O(N)$), we can replace $\sqrt{2S-b^\dagger b}$ in Eq.~(\ref{aeq:HPrep}) by $\sqrt{N}$. Also, we may safely use the expression Eq.~(\ref{eq:lindM_herm}) for $\lind_M^{(q')}$, since the sub-leading order in $N$ will only give c-number contribution. Thus, we find
\begin{equation}
    \lind_M^{(q')}
    \propto - \frac{2}{q'}(d+1)N + \left[ 2(d+1)b^\dagger b - (q'-1)(d-1) \left( b^{\dagger 2} + b^2 \right) \right] + O(N^{-1}) .
\end{equation}
Dropping the constants, we have
\begin{equation}
    \lind_M^{(q')}
    \propto \left[ 2(d+1)b^\dagger b - (q'-1)(d-1) \left( b^{\dagger 2} + b^2 \right) \right] + O(N^{-1}) .
\end{equation}
This expression is more clearly seen as the Hamiltonian of a harmonic oscillator, by introducing $\Phi = (b+b^\dagger)/\sqrt{2}$ and $\Pi = (b+b^\dagger)/(\sqrt{2}i)$, which are conjugate ``position'' and ``momentum'' operators, respectively. They satisfy $[ \Phi, \Pi ] = i$. The expression of the Lindbladian becomes
\begin{equation}    \label{aeq:harmonic}
    \lind_M^{(q')} = (q'-1)(d-1) \Pi^2 + \left( q' - (q'-2)d \right) \Phi^2    .
\end{equation}

We immediately see this low-energy effective Lindbladian is consistent with our reasoning in Sec.~\ref{ssec:MIS} from multiple aspects. The kinetic energy ($\propto \Pi^2$) vanishes in the trivial cases of $q'=1$ or $d=1$. This makes sense because if $q'=1$, single-body measurements would kill entanglement without any survival; and if $d=1$, the total Hilbert space would be trivial. On the other hand, the coefficient of the potential term ($\propto \Phi^2$) is consistent with the critical qudit dimension formula
\begin{equation}
    d_c = \frac{q'}{q'-2}, 
\end{equation}
which we have derived in Sec.~\ref{ssec:MIS} using the classical spontaneous symmetry breaking. This entanglement phase transition is in the universality class of Ginzburg-Landau $\phi^4$ transition. The sign of the potential term dictates which phase the system is in. Interestingly, when $q'=2$ the potential term does not grow with $\Phi$. This is actually a way to see why an exact $\mathrm{U}(1)$ symmetry emerges only when $q'=2$, $d=\infty$, as shown in Fig.~\ref{fig:largeS landscape}(d). In fact, as $d$ increases, the kinetic energy linearly increases while the potential is unchanged, so the excitation mode is asymptotically ``free'' as $d\to\infty$. Exactly the same situation happens when $q'>2$, $d=d_c$. The potential term also vanishes and we are left with a free quadratic kinetic energy term. 

Finally, we discuss the Liouvillian gap. It is readily seen from the harmonic oscillator form of Eq.~(\ref{aeq:harmonic}) that the low-energy spectrum has equal spacing, and the Liouvillian gap is finite, as long as $d>d_c = q'/(q'-2)$. When $q'\geq 2$ and $d=d_c=q'/(q'-2)$, the low-energy spectrum resembles that of a free particle. The Holstein-Primakoff transformation sets a length scale for the ``box size'', which is the scale where the approximation $\sqrt{2S-b^\dagger b} \approx \sqrt{2S}$ breaks down: $l \sim b \sim \sqrt{b^\dagger b} \sim \sqrt{N}$. Correspondingly, $\Delta \sim \Pi^2 \sim (1/l)^2 \sim 1/N$. Therefore, there is a different Liouvillian gap scaling between the normal purified phase ($\propto 1$) and the critical purification ($\propto N^{-1}$).

\section{Derivation of Lindbladian on two clusters}
\label{Appendix:two cluster lindblad}

In this section, we will derive the Lindbladian that governs the evolution of purities, in the case of joint random evolution or random measurement between two clusters. These terms in the Lindbladian was introduced in Sec.~\ref{ssec:2interaction} and listed in Table \ref{table:summary}. Due to the similar construction comparing to the single-cluster case, we will find little difference between this derivation and the one in the Appendix. 

We start from the $\lind_U^{(q,q)}$ Lindbladian of inter-cluster Brownian evolution. We need to calculate the average of the quadratic term in $H$, as in Eq.~(\ref{aeq:lind_average}), and the result is 
\begin{equation}
    \begin{aligned}
        \lind_U^{(q,q)} &= \frac{\mathcal{J}}{2} \sum_{\mathbf{i}, \mathbf{j}, \mathbf{a}, \mathbf{b}} \Big( T_{i_1 a_1}^{1} \cdots T_{i_q a_q}^{1} T_{j_1 b_1}^{1} \cdots T_{j_q b_q}^{1}
        - T_{i_1 a_1}^{\bar{1}*} \cdots T_{i_q a_q}^{\bar{1}*} T_{j_1 b_1}^{\bar{1}*} \cdots T_{j_q b_q}^{\bar{1}*} \\
        &\phantom{=}+ T_{i_1 a_1}^{2} \cdots T_{i_q a_q}^{2} T_{j_1 b_1}^{2} \cdots T_{j_q b_q}^{2} 
        - T_{i_1 a_1}^{\bar{2}*} \cdots T_{i_q a_q}^{\bar{2}*} T_{j_1 b_1}^{\bar{2}*} \cdots T_{j_q b_q}^{\bar{2}*} \Big)^2 \\
        &= 2\mathcal{J} \sum_{\mathbf{i}, \mathbf{j}} \Big( d^2 \cdots d^2 d^2 \cdots d^2 + dA_{i_1} \cdots dA_{i_q} dA_{j_1} \cdots dA_{j_q}   \\
        &\phantom{=}- dB_{i_1} \cdots dB_{i_q} dB_{j_1} \cdots dB_{j_q} - dC_{i_1} \cdots dC_{i_q} dC_{j_1} \cdots dC_{j_q} \Big)   ,
    \end{aligned}
\end{equation}
where $\mathbf{i} \in \{(i_1,\dots,i_q)| 1\leq i_1<\dots<i_q \leq N \}$, $\mathbf{a} \in \{(a_1,\dots,a_q)|a=1,\dots,d^2-1\}$ are indices in the first cluster, and $\mathbf{j}$, $\mathbf{b}$ similarly for the second. 

To proceed, we need to notice a convenient summation formula for the on-site matrices. For instance, for $A$ and $B$ we have
\begin{equation}    \label{aeq:leading_order}
    \begin{aligned}
    \sum\limits_{i_1<\dots<i_q}\sum\limits_{j_1<\dots<j_q}A_{i_1}\cdots A_{i_q}\cdot B_{j_1}\cdots B_{j_q}&\approx(q!)^{-1}\left(\sum_{i}A_i\right)^q\cdot(q!)^{-1}\left(\sum\limits_{j}B_i\right)^q\\&=(q!)^{-1}(q!)^{-1}(A_{\mathrm{tot}})^q(B_{\mathrm{tot}})^q  .
    \end{aligned}
\end{equation}
While this is only the leading term in the large-$N$ limit for general $q\geq 2$, it is exact when $q=1$. As a result, the final expression of the Hermitianized Lindbladian is
\begin{equation}    \label{aeq:lindU_2cluster}
    \begin{aligned}
    \lind_U^{(q,q)} \approx \frac{2\mathcal{J} d^{2q}}{(q!)^2} &\Big[ (dN)^{2q} + (2S_1^x)^{q}(2S_2^x)^{q}\\
    &-\left(\frac{d}{2}N+S_1^{x}+\sqrt{d^2-1}S_1^z\right)^{q}\left(\frac{d}{2}N+S_2^{x}+\sqrt{d^2-1}S_2^z\right)^{q}\\
    &-\left(\frac{d}{2}N+S_1^{x}-\sqrt{d^2-1}S_1^z\right)^{q}\left(\frac{d}{2}N+S_2^{x}-\sqrt{d^2-1}S_2^z\right)^{q} \Big]  .
    \end{aligned}
\end{equation}
Similarly, the Hermitianized measurement term in the large-$N$ limit reads
\begin{equation}    \label{aeq:lindM_2cluster}
    \begin{aligned}
        \lind_M^{(q',q')}
        &\approx \frac{\lambda}{(q'!)^2 (d^{2q'}+1)^2} \Big[ (d^{2q'}+1)^2 N^{2q'}  \\
        &\phantom{=} - \left( \frac{N}{2} + \sqrt{1-d^{-2}} S^z_1 + d^{-1} S^x_1 \right)^{q'} \left( \frac{N}{2} + \sqrt{1-d^{-2}} S^z_2 + d^{-1} S^x_2 \right)^{q'} \\
        &\phantom{=} - \left( \frac{N}{2} - \sqrt{1-d^{-2}}S^z_1 + d^{-1}S^x_1 \right)^{q'} \left( \frac{N}{2} - \sqrt{1-d^{-2}}S^z_2 + d^{-1}S^x_2 \right)^{q'}  \\
        &\phantom{=} - \left( S^x_1 - i\sqrt{1-d^{-2}} S^y_1 + d^{-1}\frac{N}{2} \right)^{q'} \left( S^x_2 - i\sqrt{1-d^{-2}} S^y_2 + d^{-1}\frac{N}{2} \right)^{q'}   \\
        &\phantom{=} - \left( S^x_1 + i\sqrt{1-d^{-2}} S^y_1 + d^{-1}\frac{N}{2} \right)^{q'} \left( S^x_2 + i\sqrt{1-d^{-2}} S^y_2 + d^{-1}\frac{N}{2} \right)^{q'} \Big]  .
    \end{aligned}
\end{equation}

Again, it is worth noticing that Eqs.~(\ref{aeq:lindU_2cluster}) and (\ref{aeq:lindM_2cluster}) are exact when $q$ or $q'$ is unity. Moreover, we expect these terms to be of order $O(N)$. Setting $\mathcal{J} \to (q!)^2 d^{-4q} N^{1-2q}$ and $\lambda \to (q'!)^2 d^{4q'} N^{1-2q'}$, we get the most concise expressions in the large-$d$ limit:
\begin{equation}
    \lind_U^{(q,q)}/N \approx - 2^{1-2q}\cdot \left[ (1+z_1)^q(1+z_2)^q+(1-z_1)^q(1-z_2)^q \right] ,
\end{equation}
\begin{equation}
    \begin{aligned}
        \lind_M^{(q',q')}/N \approx - 2^{-2q'}\cdot[&(1+z_1)^{q'} (1+z_2)^{q'} + (x_1+iy_1)^{q'} (x_2+iy_2)^{q'} \\
        +&(1-z_1)^{q'} (1-z_2)^{q'} +(x_1-iy_1)^{q'} (x_2-iy_2)^{q'} ] .
    \end{aligned}
\end{equation}
The real-time evolution of the purity vector is given by
\begin{equation}
    P(t) = \mathcal{N}^{-1} e^{\chi (S^x_1+S^x_2)} e^{-\lind t} e^{-\chi (S^x_1+S^x_2)} \mathcal{N} P(0)    ,
\end{equation}
where $\mathcal{N} = \mathcal{N}_1 \otimes \mathcal{N}_2$, $(\mathcal{N}_1)_{mn} = (\mathcal{N}_2)_{mn} = \sqrt{C^n_N} \delta_{mn}$, and $\tanh \chi = d^{-1}$.

\section{Phase diagram of two coupled clusters}
\label{appendix:example of two cluster phase diagram}
\begin{figure}[t]
    \centering
    \includegraphics[width=0.32\textwidth]{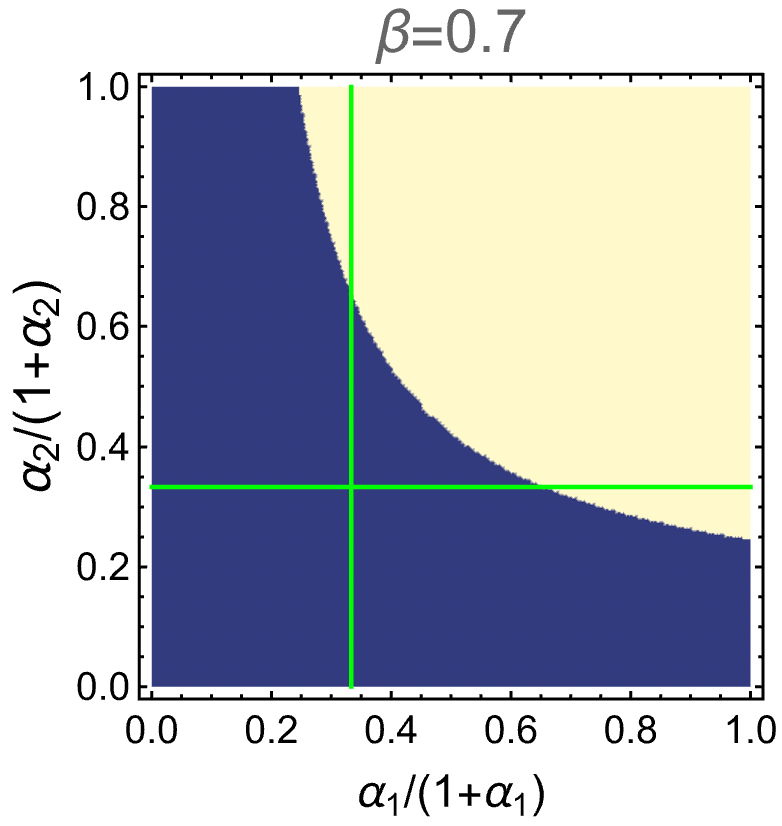}
    \includegraphics[width=0.32\textwidth]{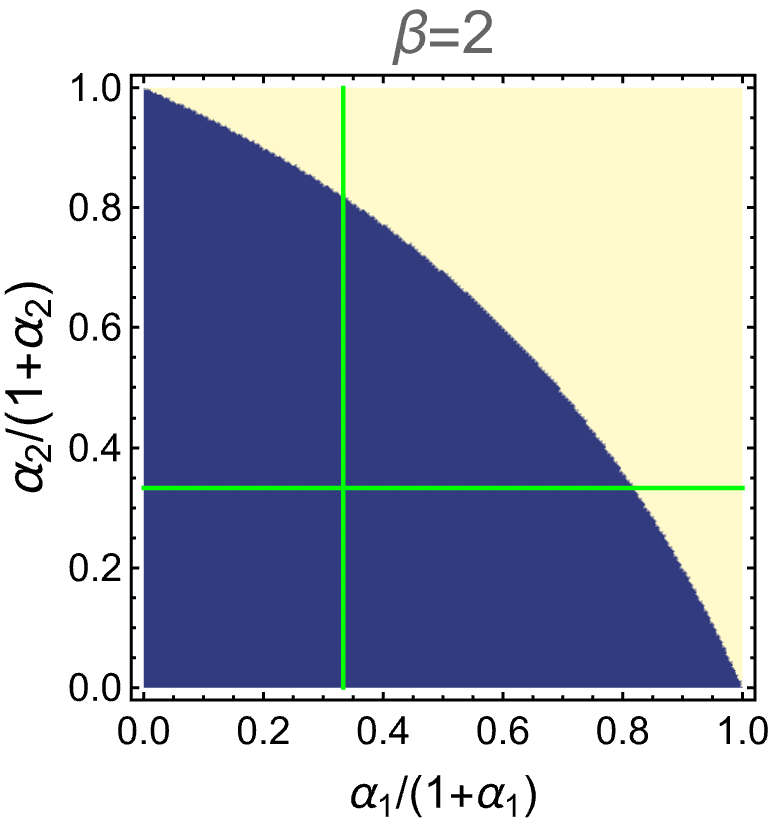}
    \includegraphics[width=0.32\textwidth]{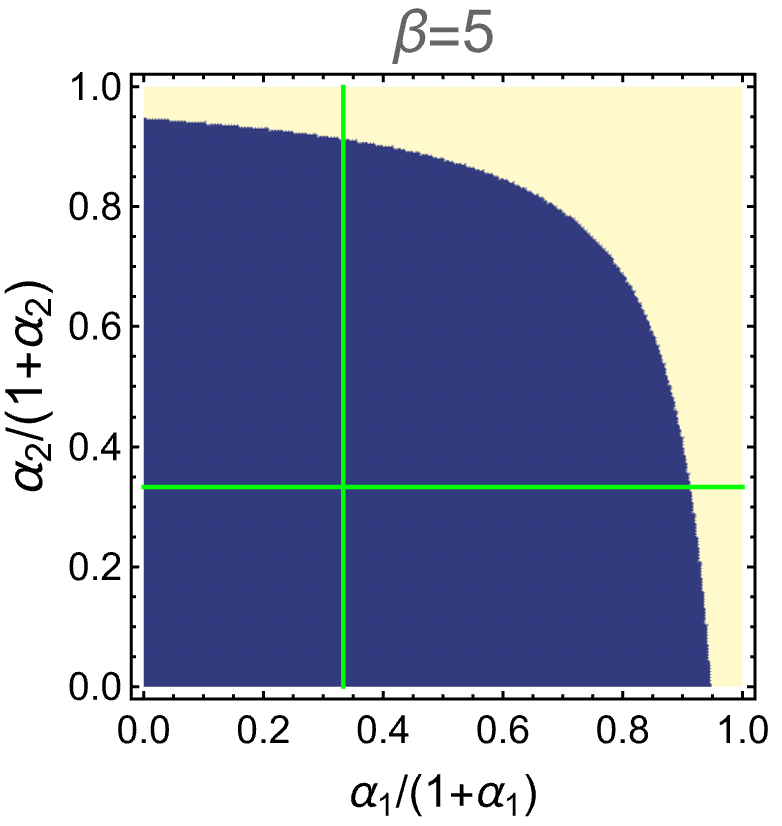}
    
    \caption{\label{fig:MIS: two cluster phase diagram} Entanglement phase diagram of the two-cluster model in Eq.~(\ref{aeq:2cluster_lind}), with qudit dimension set to be $d=2$, and for three representative choices of interaction parameter $\beta$. The horizontal/vertical axes are $\frac{\alpha_1}{1+\alpha_1},\frac{\alpha_2}{1+\alpha_2}\in[0,1]$. The yellow region is purified phase, where the ground state degeneracy is one; and the blue region is the scrambling phase, with ground state degeneracy two.   The green line is a reference for the phase diagram when interaction is turned off ($\beta=0$). \textbf{Left:} Weak interaction, $0<\beta<2$. \textbf{Middle:} Critical interaction, $\beta=2$. \textbf{Right:} Strong interaction, $\beta>2$.}
\end{figure}
To show the power of the formulation in Appendix \ref{Appendix:two cluster lindblad}, we calculate the entanglement phases under the following three effects:
\begin{enumerate}
    \item Intra-cluster two-body random unitaries, the strength being $\mathcal{J}_1 = \mathcal{J}_2 = d^{-2} N^{-1}$;
    \item Intra-cluster single-body measurement with different strength in two clusters: $\lambda_1 = 2d^2(d+1)\alpha_1$, $\lambda_2 = 2d^2(d+1)\alpha_2$;
    \item Inter-cluster $(1,1)$-body random unitaries: $\mathcal{J}_{12} = \beta d^{-2} N^{-1}$. 
\end{enumerate}
Working in the large-$N$ limit, the total Lindbladian is
\begin{equation}    \label{aeq:2cluster_lind}
    \begin{aligned}
    \lind/N &= \left( \lind_U^{(2,0)} + \lind_U^{(0,2)} + \lind_M^{(1,0)} + \lind_M^{(0,1)} + \lind_U^{(1,1)} \right)/N \\
    &= \frac{1}{2}(x_1^2+x_2^2+2\beta x_1x_2)-\frac{1}{2}(d^2-1)(z_1^2+z_2^2+2\beta z_1z_2)\\
    &\phantom{=} - (\beta+1)d(x_1+x_2)-2d(\alpha_1x_1+\alpha_2x_2)  ,
    \end{aligned}
\end{equation}
where the superscripts of $\lind_{U/M}$ label the number of qudits that an operation involves. 

Generally, the entanglement phases can be recognized by minimizing the classical problem Eq.~(\ref{aeq:2cluster_lind}). In the current two-cluster problem, recognizing the entanglement phases still amounts to counting the ground-state degeneracy. Knowing that the minima require $y_1=y_2=0$, we may hence parametrize the classical spin as $\lind(x_1=\cos{\varphi_1},z_1=\sin{\varphi_1},x_2=\cos{\varphi_2},z_2=\sin{\varphi_2})$. In the unbroken phase the minimum lies at $\varphi_1=\varphi_2=0$, so the problem reduces to observing the second-order derivative (eigenvalues of the Hessian matrix) at $\varphi_1=\varphi_2=0$. 

In Fig.~\ref{fig:MIS: two cluster phase diagram}, we show a complete entanglement phase diagram of this model Eq.~(\ref{aeq:2cluster_lind}) with variable $\alpha_1$, $\alpha_2$ and $\beta$. According to different interaction strength $\beta$, the model have qualitatively differently:
\begin{enumerate}
\item[(i).]\textit{No interaction, }with $\beta=0$. Two clusters are decoupled. Whether an individual cluster is in purified/scrambled phase depends on their own measurement strength $\alpha_{1,2}$ smaller or greater than $1/2$. For any $\beta>0$, the two clusters are either both purified or both scrambled.
\item[(ii).]\textit{Weak interaction, }with $0<\beta<2$. In this regime, the `connectivity' between two clusters is low, in the sense that if one only performs measurement on cluster 1, one can never purify the whole system (when $\alpha_2=0$, even tuning $\alpha_1=+\infty$ is still in scrambling phase). In fact, one needs to at least perform a minimal amount measurement on cluster 2 ($\alpha_2>\frac{2-\beta}{4}$), such that one can increase $\alpha_1$ beyond some finite value to purify the system.
\item[(iii).]\textit{Critical interaction, }with $\beta=2$. This is the phase boundary between (ii) and (iv). One also notice that here we have $\mathcal{J}_1=\mathcal{J}_2=\frac{1}{2}\mathcal{J}_{12}$, so the permutation symmetry between qudits are enhanced from $\text{perm}(N)\times\text{perm}(N)\times\mathbb{Z}_2$ to $\text{perm}(2N)$.
\item[(iv).]\textit{Strong interaction, }with $\beta>2$. In this regime, the `connectivity' between two clusters is high. Comparing to (ii), one can actually perform measurement on cluster 1 only in order to purify the whole system ($\alpha_2=0,\alpha_1>\frac{(2\beta-1)(\beta+1)}{\beta-2}$ is in purified phase).
\end{enumerate}
From the discussion above, one see that the inter-cluster interaction strength affect the purification efficiency (whether one can purify the whole system by measuring only one cluster).

\section{Luttinger liquid explanation of the entropy in SU(2) symmetric model}
\label{app: Luttinger}
In section~\ref{ssec:su2}, we observed that for a 1D chain with nearest-neighbor two-body measurement, starting from a local product state, the entropy of a single cut region with length $\l$ in the late time steady state satisfy a $\text{CFT}_2$ behaviour $S_{\text{PBC}}(l)=\frac{1}{2}\log|l|$. We will explain this using Luttinger liquid theory description of low energy excitations of XXZ model. We will mainly follow~\cite{PhysRevB.46.10866,Affleck:1988zj}.

We first translate procedure of purity calculation to a correlation function in the spin chain model. Consider the following spin-1/2 Hamiltonian on 1d lattice:
\begin{equation}
\mathcal{L}=\sum_i -S^x_iS^x_{i+1}-S^z_iS^z_{i+1}+S^y_iS^y_{i+1}
\end{equation}
where $S^{x,y,z}_i:=\frac{1}{2}\sigma_i^{x,y,z}$ are spin-1/2 operators. Consider initial state:
\begin{equation}
|\psi_0\rangle :=\bigotimes_{i}|+\hat{x}\rangle_i
\end{equation}
and left boundary condition:
\begin{equation}
|X_A\rangle:=\bigotimes_{i\in A}|-\hat{z}\rangle_i\bigotimes_{i\notin A}|+\hat{z}\rangle_i
\end{equation}
Then the steady state purity and second Renyi entropy is defined by 
\begin{equation}
P_A\propto\langle X_A|e^{-t \mathcal{L}}|\psi_0\rangle,\ S_A=-\log P_A,\ t\rightarrow\infty
\end{equation}

The standard manipulation is to utilize Jordan-Wigner transformation to go from spin language to fermion language:
\begin{equation}
n_i:=f_i^\dagger f_i=\frac{1}{2}(1-2S^y_i)
\end{equation}
\begin{equation}
S^+_i:=S^z_i+iS^x_i=f_i\prod_{j<i}(2S^y_i)=f_i\exp(i\pi\sum_{j<i}n_i)
\end{equation}
where $f_i$ satisfies $\{f_i^\dagger,f_j\}=\delta_{ij}$.
Now in the fermion language, the Hamiltonian becomes:
\begin{equation}
\mathcal{L}=-\frac{1}{2}\sum_i(f_{i+1}^\dagger f_i+f_{i}^\dagger f_{i+1})+\sum_i(n_{i+1}-\frac{1}{2})(n_{i}-\frac{1}{2})
\end{equation}
First ignoring the density-density interaction, the hopping terms results in a single partitle spectrum $\varepsilon(k)=-\cos k$. The global (for all charge sector) ground state is in half-filling sector, with two Fermi point at $\pm k_F$ with Fermi momentum being $k_F=\frac{\pi}{2}$.

In low energy limit, the particle-hole excitation near two Fermi points has well defined gapless linear dispersion relation with quadratic action after bosonization. It turns out that action remains quadratic in boson language even after including density-density interaction, if Umklapp back-scattering is ignored.

Upon bosonization, these density fluctuation can be described by two bosonic fields $\varphi\text{ and }\theta$, which are related to field theory variable by:
\begin{equation}
n_i\simeq\rho(x)=\frac{1}{2\pi}\partial_x\varphi+\bar{\rho}
\end{equation}
\begin{equation}
S^+_i\simeq e^{i\theta(x)/2}
\end{equation}
where $\bar{\rho}$ is the uniform average density in ground state.

The effective description of density fluctuation in terms of the real boson field $\varphi(\tau,x)$ is given by the following Euclidean action for Luttinger liquid:
\begin{equation}
I[\varphi]=\frac{K}{8\pi}\int d\tau dx\cdot (\partial_\tau\varphi)^2+(\partial_x\varphi)^2,\ \  \mathcal{Z}=\int D\varphi\  e^{-I[\varphi]}
\end{equation}
where $K$ is the Luttinger parameter of XXZ chain. For our case of SU(2) symmetric XXZ chain, we have $K=2$. To recover $\theta$ field, one first notice that this action can be equivalently obtained by integrating out conjugate momentum $\Pi(\tau,x)$:
\begin{equation}
I[\varphi,\Pi]=\int d\tau dx \cdot \frac{K}{8\pi}(\partial_x\varphi)^2+i\Pi\partial_\tau\varphi+\frac{2\pi}{K}\Pi^2,\ \ \mathcal{Z}=\int D\Pi D\varphi\  e^{-I[\varphi,\Pi]}
\end{equation}
In Luttinger liquid, we have $\Pi(\tau,x):=-\frac{1}{4\pi}\partial_x\theta(\tau,x)$, so the action in terms of $\varphi,\theta$ is:
\begin{equation}
I[\varphi,\theta]=\frac{1}{8\pi}\int d\tau dx \cdot K(\partial_x\varphi)^2-2i\partial_x\theta\partial_\tau\varphi+K^{-1}(\partial_x\theta)^2
\end{equation}
Now, one define:
\begin{equation}
\varphi:=\frac{1}{\sqrt{K}}(\varphi_L+\varphi_R),\ \theta:={\sqrt{K}}(\varphi_L-\varphi_R)
\end{equation}
We find that up to total derivative terms, the real boson $\varphi_L,\varphi_R$ decoupled:
\begin{equation}
I[\varphi_L,\varphi_R]=\frac{1}{4\pi}\int d\tau dx\cdot (\partial_x\varphi_L)(\partial_x-i\partial_\tau)\varphi_L+\frac{1}{4\pi}\int d\tau dx\cdot (\partial_x\varphi_R)(\partial_x+i\partial_\tau)\varphi_R
\end{equation}
In Euclidean signature we can define $z=x+i\tau$, one find that the e.o.m. of $\varphi_L$ is:
\begin{equation}
\frac{1}{2}(\partial_x-i\partial_\tau)\partial_x\varphi_L\equiv \partial_z\partial_x\varphi_L=0
\end{equation}
So, we find that $\partial_x\varphi_L$ is holomorphic. The field $\varphi_{L/R}$ are chiral boson, and this is called chiral Luttinger liquid. A standard calculation shows that the equal imaginary time two-point correlator of vertex function is given by:
\begin{equation}
\langle e^{i\lambda\varphi_L(0,x_1)}e^{-i\lambda\varphi_L(0,x_1)}\rangle=\frac{1}{|\Lambda(x_1-x_2)|^{\lambda^2}}
\end{equation}
where $\Lambda$ is a UV cutoff parameter.

Now, we are ready to attack on our problem of calculating purity. We first work out the boundary condition changing operator. Using $e^{i\pi S^y}|+\hat{z}\rangle=|-\hat{z}\rangle$, we have:
\begin{equation}
|X_A\rangle=\exp(i\pi\sum_{j\in A}S_j^y)\bigotimes_i|+\hat{z}\rangle_i:=\exp(i\pi\sum_{j\in A}S_j^y)|\Omega\rangle
\end{equation}
Recall that we perform Jordan Wigner transformation in $x$ basis: $n_i=\frac{1}{2}(1-2S^y_i)$. We consider single interval region $A$, where the left/right end point in continuous coordinate being $x_1,x_2$:
\begin{equation}
\sum_{j\in A}n_j\simeq\int_{x_1}^{x_2}dx\ \rho(x)=\int_{x_1}^{x_2}dx(\frac{1}{2\pi}\partial_x\varphi+\bar{\rho})=\frac{1}{2\pi}\varphi(x_2)-\frac{1}{2\pi}\varphi(x_1)+\bar{\rho}(x_2-x_1)
\end{equation}
So, we find:
\begin{equation}
\begin{aligned}
P_A&\propto\langle{X_A|e^{-\infty \mathcal{L}}|\psi_0}\rangle\\
&\propto e^{-\frac{1}{2}i\pi\sum_{j\in A}1}\langle{\Omega|e^{i\pi\sum_{i\in A}n_j}|G}\rangle\\
&\propto\langle{\Omega|e^{i\varphi(x_2)/2}e^{-i\varphi(x_1)/2}|G}\rangle
\label{eq: PA luttinger}
\end{aligned}
\end{equation}

In the second line, we drop the phase factor $e^{-\frac{1}{2}i\pi\sum_{j\in A}1}$ since we know the final result is real and positive, and we also define $|G\rangle\propto e^{-\infty H}|\psi_0\rangle$ being the ground state among all charge sector. In fact we know precisely that this global ground state $|G\rangle$ is in the sector where $S^x_{\text{total}}=0$. In fermion language, this is the half-filling sector. One can see this by either appealing to exact result about spin chain, or pertubatively treating fermion density interaction term $\sum_iS^y_iS^y_{i+1}$ being small. Then the first two terms $\sum_i-S^x_iS^x_{i+1}-S^z_iS^z_{i+1}$ will give $\varepsilon(k)=-2\cos k$ single particle dispersion, which has minimal energy in half-filling. One then can easily  check that $|\psi_0\rangle$ has non-trivial projection onto  sector.

Notice that this form of $P_A$ is not a usual 2pt yet, because of a fixed boundary condition $\langle\Omega|$. One first want to rewrite the boundary changing operator solely in chiral boson $\varphi_L$.

The trick is to notice that in calculating $P_A$ in~\eqref{eq: PA luttinger}, we can multiply any number to r.h.s. as long as it doesn't depend on choice of $A$. 

Notice that
schematically, we have $S^z_i\simeq\cos(\frac{\theta(x)}{2})$. So the boundary state $|\Omega\rangle$ which corresponds to the uniformly maximally polarized state in $+\hat{z}$ direction is equivalent to set the boundary condition to be $\theta(x)=0,\forall x$ in continuous theory.

This is equivalent to set $\varphi_L(x)=\varphi_R(x),\forall x$ at boundary. Using this boundary condition, we can first rewrite the boundary changing operator solely in $\varphi$:
\begin{equation}
P_A\propto \langle\Omega|e^{i\varphi_L(x_2)/\sqrt{K}}e^{-i\varphi_L(x_1)/\sqrt{K}}|G\rangle
\end{equation}
To evaluate $P_A$, we first represent $|G\rangle$ as a path integral of non-chiral theory on the lower half plane ($\tau<0,x\in\mathbb{R}$). Then we notice that the Dirichlet boundary condition $\varphi_L(x)=\varphi_R(x)$ is a reflective wall at $\tau=0$, so we can move the degree of freedom of $\varphi_R$ on lower half plane to $\varphi_L$ on the upper half plane:
\begin{equation}
\begin{aligned}
P_A&\propto\int_{\tau<0,x\in\mathbb{R},\varphi_L(0,x)=\varphi_R(0,x)} D\varphi_LD\varphi_R \cdot e^{i\varphi_L(0,x_2)/\sqrt{K}}e^{-i\varphi_L(0,x_1)/\sqrt{K}}e^{-I[\varphi_L]-I[\varphi_R]}\\
&=\int_{\tau\in\mathbb{R},x\in\mathbb{R}} D\varphi_L\cdot e^{i\varphi_L(0,x_2)/\sqrt{K}}e^{-i\varphi_L(0,x_1)/\sqrt{K}}e^{-I[\varphi_L]}\\
&\propto\langle{e^{i\varphi_L(0,x_2)/\sqrt{K}}e^{-i\varphi_L(0,x_1)/\sqrt{K}}}\rangle_{\varphi_L \text{ on }\mathbb{R}^2}\\
&\propto \frac{1}{|\Lambda(x_1-x_2)|^{\frac{1}{2}}}
\end{aligned}
\end{equation}
In the last line, we plug in $K=2$ for SU(2) symmetric XXZ model. Finally, we obtain the entropy being:
\begin{equation}
S_A=-\log P_A=\frac{1}{2}\log|x_1-x_2|
\end{equation}
The coefficient $1/2$ exactly matches the entropy curve obtained numerically in Fig. \ref{fig:vertex illustration and DMRG} (b).

\bibliographystyle{JHEP}
\bibliography{references}

\end{document}